\documentclass[prd,twocolumn,superscriptaddress,floatfix,amsmath,amssymb]{revtex4-1}
\usepackage[english]{babel}
\usepackage{graphicx}
\usepackage{verbatim}
\usepackage{mathrsfs}
\usepackage{cancel}
\usepackage{bm}
\usepackage{color} 

\newcommand{\ket}[1]{\left| {#1} \right\rangle}
\newcommand{\bra}[1]{\left\langle {#1} \right|}

\newcommand{\partialfrac}[2]{\frac{\partial #1}{\partial #2}}
\newcommand{\proj}[2]{\left| {#1} \right\rangle\!\left\langle {#2} \right|}
\newcommand{\ii}{\mathrm{i}}

\begin{document}
\title{Casimir forces on atoms in optical cavities}
\author{\'{A}lvaro M. Alhambra}
\affiliation{Department of Applied Mathematics, University of Waterloo, Waterloo, Ontario, N2L 3G1, Canada}
\affiliation{Institute for Quantum Computing, University of Waterloo,  Waterloo, Ontario, N2L 3G1, Canada}
\author{Achim Kempf}
\affiliation{Department of Applied Mathematics, University of Waterloo, Waterloo, Ontario, N2L 3G1, Canada}
\affiliation{Institute for Quantum Computing, University of Waterloo, Waterloo, Ontario, N2L 3G1, Canada}
\affiliation{Perimeter Institute for Theoretical Physics, Waterloo, Ontario, N2L 2Y5, Canada}
\author{Eduardo Mart\'{i}n-Mart\'{i}nez}
\affiliation{Department of Applied Mathematics, University of Waterloo, Waterloo, Ontario, N2L 3G1, Canada}
\affiliation{Institute for Quantum Computing, University of Waterloo, Waterloo, Ontario, N2L 3G1, Canada}
\affiliation{Perimeter Institute for Theoretical Physics, Waterloo, Ontario, N2L 2Y5, Canada}

\begin{abstract}
Casimir-type forces, such as those between two neutral conducting plates, or between a sphere, atom or molecule and a plate have been widely studied and are becoming of increasing significance, for example, in nanotechnology. A key challenge is to better understand, from a fundamental microscopic approach, why the Casimir force is in some circumstances attractive and in others repulsive. Here, we study the Casimir-Polder forces experienced by small quantum systems such as atoms or molecules in an optical cavity. In order to make the problem more tractable, we work in a 1+1 dimensional setting, we take into account only the ground state and first excited state of the atom and we model the electromagnetic field as a scalar field with Dirichlet or Neumann boundary conditions. This allows us to determine the conditions for the Casimir force to be attractive or repulsive for individual atoms, namely through the interplay of paramagnetic and diamagnetic vacuum effects. We also study the microscopic-macroscopic transition, finding that as the number of atoms in the cavity is increased, the atoms start to affect the Casimir force exerted on the cavity walls similarly to a dielectric medium. 
\end{abstract}

\maketitle  

\section{Introduction}

The presence of Casimir forces between a neutral atom and a conducting plate was noticed already in the early days of quantum field theory \cite{CasimirPolder}. They were first detected experimentally in \cite{Sukenik}. The analysis of Casimir-type forces has been extended to various different geometries and for both conductive and dielectric media \cite{Marvin,Buhmann}. It was found that Casimir forces are in general attractive. However, evidence for instances in which the Casimir force is repulsive have also been found (see \cite{Lifshitz,NewDevelopments}). While the conditions for when the Casimir force becomes repulsive are not yet fully understood, several proposals for setups which should yield repulsive Casimir forces have been made. These include setups with conditions of high magnetic permeability \cite{Boyer,Kenneth}, optical setups involving left-handed metamaterials \cite{Leonhardt}, or non--trivial boundary conditions \cite{Levin}. Examples of repulsive Casimir forces have been experimentally observed only in recent years \cite{Munday}. Repulsive Casimir forces are of practical interest, for example, in the field of nanomachines \cite{ChanScience}, for which Casimir-related effects are the ultimate source of friction. See \cite{ReviewPelotas} for a detailed review.

Here, motivated by progress in quantum optics and its applications to quantum information (see among many others \cite{Knill:2001qf,Robort,Reznik1,Harvesting}), we will study the Casimir effect experienced by atoms or molecules inside optical cavities. In particular, we will study the conditions for the Casimir force to become repulsive. 

Technically, Casimir-type forces arise when the ground state energy $E_0(\lambda)$ of a quantum system depends on a classically-treated parameter, $\lambda$, where $\lambda$ is usually the distance between two subsystems. Then, $F=dE_0/d\lambda$ is the Casimir force with which the quantum system drives or resists an adiabatically slow change of $\lambda$.   
Ultimately, the Casimir-Polder forces between neutral systems such as neutral atoms or neutral macroscopic matter arises via the standard QED minimal coupling between the constituent charges and the electromagnetic field \cite{CasimirPolder}:
\begin{equation}\label{eq:QED}
H^{\text{QED}} =\frac{1}{{2m}} \left[\bm p -\frac{e}{ c} {\bm{A}}({\bm{x}})\right]^2
\end{equation}
In the Coulomb gauge, $\sum_{i=1}^3\partial_i \bm A^i(\bm x)=0$, the interaction terms read as 
\begin{equation}\label{eq:QED2}
H_{\text{I}}^{\text{QED}} = -\frac{e}{m c}  {\bm{A}}({\bm{x}}) \cdot {\bm{p}}  +\frac{e^2}{2 m c^2}[\bm{A}({\bm{x}})]^2
\end{equation}
In weak  ${\bm{A}}$  fields, the term linear in $\bm{A}$, the so-called paramagnetic term, usually dominates over the term which is quadratic in $\bm{A}$, the so-called diamagnetic term. The diamagnetic term is therefore often neglected in simplified models of light--matter interaction. The Casimir effect between neutral systems would seem to be a small-field case in which the diamagnetic term is negligible. 
This is because in this case there is no finite classical background field $\bm{A}$ and the neutral systems in question merely interact with the quantum fluctuations of $\bm{A}$ in the vacuum. The variance, $\langle \bm{A}^2 \rangle$,  of the quantum fluctuations of $\bm{A}$ in the vacuum depends however, on the size of the region that they are being smeared over, i.e., the variance depends on the size of the system that interacts with the $\bm{A}$ field. As the size of the system is taken to zero the variance of the quantum fluctuations of $\bm{A}$ diverges. For small enough systems, therefore, the diamagnetic term, which is proportional to the variance, can contribute on the order of the paramagnetic term, as we will see in the following. 

In order to make the calculations more tractable, we will use a simplified model, the so-called Unruh-DeWitt (UdW) model. In the UdW model, the interaction between an atom and a quantum field is described as the interaction of a localized two--level system with a scalar field. The UdW model has been shown to reproduce very well the light-matter interaction at leading order, as long as there is no exchange of orbital angular momentum \cite{Wavepackets}. 

Concretely, we will begin by analyzing the Casimir-Polder force experienced by a neutral atom between reflective plates. For previous work in this field, see e.g.,  \cite{Rizzuto,RizzutoDynamical,RizzutoPassante,RizzutoSpagnolo}. Since we are here working with the UdW model, we can go beyond the usual proximity field approximation. Namely, instead of considering one atom close to a mirror \cite{CasimirPolder}, we allow atoms to be introduced at any arbitrary distances from the two mirrors of a cavity.  Also, we will be able to add the diamagnetic term and conclude that the diamagnetic term should generally be taken into account. Finally, we will explain both quantitatively and intuitively how the paramagnetic and diamagnetic terms together with the boundary conditions in the cavity determine whether the Casimir force is repulsive or attractive.

We will then also determine how the presence of multiple atoms in the optical cavity affects the Casimir-type force between the walls of the cavity. Namely, in the regime of a high density of uniformly-distributed atoms, the collection of atoms starts to act as a dielectric medium so that in a suitable approximation the macroscopic Lifshitz formalism \cite{Lifshitz} for the calculation of the Casimir effect between plates \cite{Casimir} separated by a dielectric medium becomes applicable. We find scenarios where the Casimir force between the plates and the forces due to the presence of the atoms oppose each other, so that in certain regimes the total force between the plates can become repulsive. This result from a microscopic description supports previous works that used macroscopic effective models, suggesting that Casimir forces can be screened in the presence of matter \cite{Raabe,Tomas}.

\section{Field-matter interaction models}

We begin by considering the standard Unruh-DeWitt Hamiltonian \cite{DeWitt} for the interaction between a scalar field and an atom modeled by a  two-level system. This model is obtained by replacing the vector quantum field ${\bm{A}}$ by a scalar quantum field $\phi$ and by reducing the Hilbert space of the atom from infinite dimensions to just the two dimensions of the ground state and first excited state. 
The paramagnetic term of the minimal coupling Hamiltonian (\ref{eq:QED2}) then takes the form:
\begin{equation}\label{preH}
H_{I}=  \lambda\, m_\text{d}\,\phi(x_\text{d}).
\end{equation}
Here, $\phi(x_\text{d})$ plays the role of ${\bm{A}}({\bm{x}})$. 
The $2\times 2$ matrix ${m}_\text{d}$ represents the action of the paramagnetic term on the two-dimensional Hilbert space of the atom which is spanned by its ground state and its first excited state:
\begin{equation}\label{eq:monop} m_\text{d}=\ket{g}\bra{e}+\ket{e}\bra{g}=\sigma_{-}+\sigma_{+},
\end{equation}
Here, $\lambda$ is the coupling strength. 

Notice that in this Hamiltonian one models the atom as coupling to the field at a point only, see, e.g., \cite{Wavepackets}, on the assumption that the spatial extent of the atom is negligible as compared to the wavelength of the radiation that is resonant with the atom's energy gap. 

Notice also that the model assumes that the atom will behave like an effective two--level system. This means that the time evolution will not induce transitions outside of the sector spanned by the two lowest energy states. This is a good approximation, for example in the case of hydrogen (neglecting spin degeneracy). There, the probability of a transition from the 1s to the 2p level is negligible compared to both, the probability of remaining in the 1s state and the already small probability of being excited to 2s. This is discussed  in a mathematically rigorous way in the paper \cite{Wavepackets}, where it is shown that indeed, the Unruh-DeWitt model does reproduce the physics of the term $\bm A(\bm x) \cdot \bm p$ in the atomic electron - electromagnetic field interaction if no exchange of orbital angular momentum is involved (the photon-carried angular momentum is balanced by the electron spin). 

 Nevertheless, it is important to keep in mind that the operator \eqref{eq:monop} is a simplification that stems from a dimensional reduction of the Hilbert space of $\bm p$, which means that one one has to carefully consider the appropriate matrix elements that need to be taken into account to model transitions between different orbitals. The associated subtleties will be analyzed in more detail in section \ref{sec:phisqrd} when we introduce the specific spatial profiles.

Following, we will also add an analog of the QED diamagnetic term to the Unruh deWitt model. Before, however, let us analyze the behavior of the paramagnetic term. 

In the case of an atom in an optical cavity, we can expand the field in terms of the well-known stationary solutions of the Klein-Gordon equation \cite{CasimirPolder}
\begin{equation}\label{eq:Hi}
H_{I} =  \lambda\,  m_\text{d} \sum_{j=1}^{\infty} [a_{j}^{\dagger}+a_{j} ]\frac{\sin{k_{j} x_\text{d}}}{\sqrt{\omega_{j}L}}
\end{equation}
for the case of a reflective cavity and
\begin{equation}\label{eq:HiVN}
H_{I} =  \lambda\,  m_\text{d} \sum_{j=1}^{\infty} [a_{j}^{\dagger}+a_{j} ]\frac{\cos{k_{j} x_\text{d}}}{\sqrt{\omega_{j}L}}
\end{equation}
for the case of a cavity whose fields obey Neumann boundary conditions (see, for instance \cite{Wilson}, for the one-dimensional case).
The fact that a scalar field is considered instead of the electromagnetic field does not, in itself, introduce any fundamental differences in the nature of the model. The electric and magnetic contributions can often be separately modeled through scalar fields obeying corresponding boundary conditions. Scalar fields have been used to analyze Casimir-type phenomena, e.g., in \cite{MiltonScalar,SchadenPRL}.  One small caveat is that this simple model encodes the basic features of the light-matter interaction for atomic transitions only in the absence of the exchange of orbital angular momentum \cite{Wavepackets}. The model has been used in studies of Casimir-Polder forces involving only one conducting plate \cite{Rizzuto,RizzutoDynamical,RizzutoPassante,RizzutoSpagnolo} and is commonly used in studies of quantum field theory in curved space-times and relativistic quantum information \cite{Crispino,MartinMartinezMenicucci,oscillator,Brusko,Barbadooor,BandD}, as well as in quantum optics (see, e.g.,  \cite{ScullyBook}).

However, the model does not contain the analog of a diamagnetic field self-interaction term as in (\ref{eq:QED}). 
To include such a term, we add to the interaction Hamiltonian: 
\begin{align}\label{eq:Hi2}
 H_{I}' =  \lambda m_\text{d}\phi(x_\text{d}) +\alpha~\frac{\lambda^2}{\Omega} [\phi(x_\text{d})]^2
\end{align}
In the new term, the squared coupling constant is divided by the atomic energy gap $\Omega$ to provide the correct units. Unlike the electromagnetic minimal coupling in (\ref{eq:QED}), the UdW model does not uniquely determine the dimensionless constant $\alpha$, i.e., it does not determine the prefactor of the diamagnetic term, except of course that $\alpha$ should be positive. 

As we will discuss in Section \ref{sec:phisqrd}, the best choice of $\alpha$ in a simple UdW model for the full electromagnetic (EM) interaction will depend on the spatial profiles of the relevant orbitals of the specific atoms considered. Since our aim in working with a simplifying UdW model is to obtain a qualitative understanding of the range of possible effects, we will first analyze the dependence of the Casimir-Polder interactions on the value of $\alpha$. Then, in order to work with a definite model, we will choose a particular value of $\alpha$ that will let us explore the regime when the diamagnetic term contributes significantly. This value of $\alpha$ is also natural in that it yields a natural dependence on the atomic gap that is similar to the one corresponding to the full electromagnetic case \cite{CasimirPolder}. 


We notice that the diamagnetic term contains only field operators, namely $\phi^2$, but no operators of the atom's quantum system. This means that the lowest order coupling of this term to the dynamics of the atom is with the third power of the coupling constant. The diamagnetic term's contribution is therefore negligible for the atom's dynamics up to second order in the perturbative expansion. We will nevertheless include this term, since as we shall show in the following sections, its presence quantitatively and qualitatively affects the Casimir-Polder forces.

To this end, we first need to address, however, the fact that with the diamagnetic $\phi^2$ term in the Hamiltonian \eqref{eq:Hi2}, the atom cannot be assumed to couple to the field at a single point only. The reason is that a field's vacuum quantum fluctuations at a point diverge: $\langle 0\vert \phi^2(x)\vert 0\rangle=\infty$. To regularize the divergence, a finite spatial profile for the atom has to be introduced,  which then ensures that the atom couples to the field's quantum fluctuations smeared over a volume, and these smeared fluctuations are finite. The original article by Casimir and Polder \cite{CasimirPolder}  \textbf{introduced a regularizing factor $e^{-\gamma k}$ as an effective spatial profile with the limit $\gamma\rightarrow0$ taken at the end. Here, we instead regularize this divergence by setting the profile of the atomic levels to be the actual radial wavefunction of atomic $s$ orbitals, as we will explain in more detail in Section \ref{sec:phisqrd}.}





\section{Single atom In a reflective cavity} \label{sec:OneD}

We begin by characterizing the force felt by single atoms in optical cavities, and the effect of their presence on the cavity walls. In particular, we will study the potential  role of the diamagnetic term in the Casimir effect. To this end, we will consider the light-matter interaction in optical cavities with and without the diamagnetic term, with various boundary conditions. We will see that the exclusion or inclusion of the diamagnetic term in the interaction Hamiltonian can make the Casimir force attractive or repulsive. We will also find a switch between atractive and repulsive Casimir forces when the boundary conditions are switched between Dirichlet and Neumann conditions.

\subsection{One atom in a cavity, without diamagnetic term} \label{sec:OneDUdW}

Our aim is to calculate the ground state energy of the system consisting of an atom in a cavity with Dirichlet boundary conditions, up to second order in perturbation theory. To this end, we will use the  interaction Hamiltonian (\ref{eq:Hi}), which models an Unruh-DeWitt type of interaction without diamagnetic term. Using time-independent perturbation theory, the leading order correction to the energy of the ground state is of second order in the coupling strength. A standard calculation yields that the energy of the ground state is given by
\[E_I=E_0 +E^{(2)}+\mathcal{O}(\lambda^4)\]
where $E_0$ is the energy of the free Hamiltonian's ground state and where the energy difference between the ground states of the free and the interacting system,
\[\delta E = E^{(2)}+\mathcal{O}(\lambda^4),\]
obeys:
\begin{align} \label{eq:secen}
\nonumber E^{(2)}=&\sum_{l=1}^{\infty}\frac{-\lambda^2}{\omega_{l}+\Omega}\Big|\bra{e,k_{l}} m_\text{d} \sum_{j=1}^{\infty} (a_{j}^{\dagger}+a_{j} )\frac{\sin{k_{j} x_\text{d}}}{\sqrt{\omega_{j}L}}
\ket{g,0}\Big|^{2} \\
=&\sum_{j=1}^{\infty}\frac{-\lambda^{2} \sin^2\left(\frac{\pi j}{L}x_{d}\right)}{(\pi j/L+\Omega)(\pi j)}
\end{align} 
We substituted the frequency of the discrete modes $\omega_{j}=\pi j/L$ and made use of the dispersion relation. The series can be summed analytically. Namely,
\begin{align}
\nonumber&E^{(2)}=\frac{-\lambda ^2}{4 \pi  \Omega } \left[2 {\cal{H}}\left({\frac{L \Omega }{\pi }}\right)+e^{\frac{2 i \pi  x_\text{d}}{L}} \Phi \left[e^{\frac{2 i \pi  x_\text{d}}{L}},1,\frac{L \Omega }{\pi }+1\right] \right. \\
\nonumber  &+ \left. e^{-\frac{2 i \pi  x_\text{d}}{L}} \Phi \left[e^{-\frac{2 i \pi  x_\text{d}}{L}},1,\frac{L \Omega }{\pi }+1\right]+\log \left(2-2 \cos \frac{2 \pi  x_\text{d}}{L}\right)\right]\\
\end{align}
where
\begin{align} \label{eq:definition}
\!\!\!\!\Phi\left[z, s,\alpha \right]=\sum_{n=0}^{\infty}\frac{z^n}{(n+\alpha)^{s}}\,\,; \quad {\cal{H}}(x)=x \sum_{k=1}^{\infty}\frac{1}{k(x+k)}
\end{align}
are respectively the Lerch transcendent and the generalized harmonic number.

\begin{figure}
\includegraphics[width=0.48\textwidth]{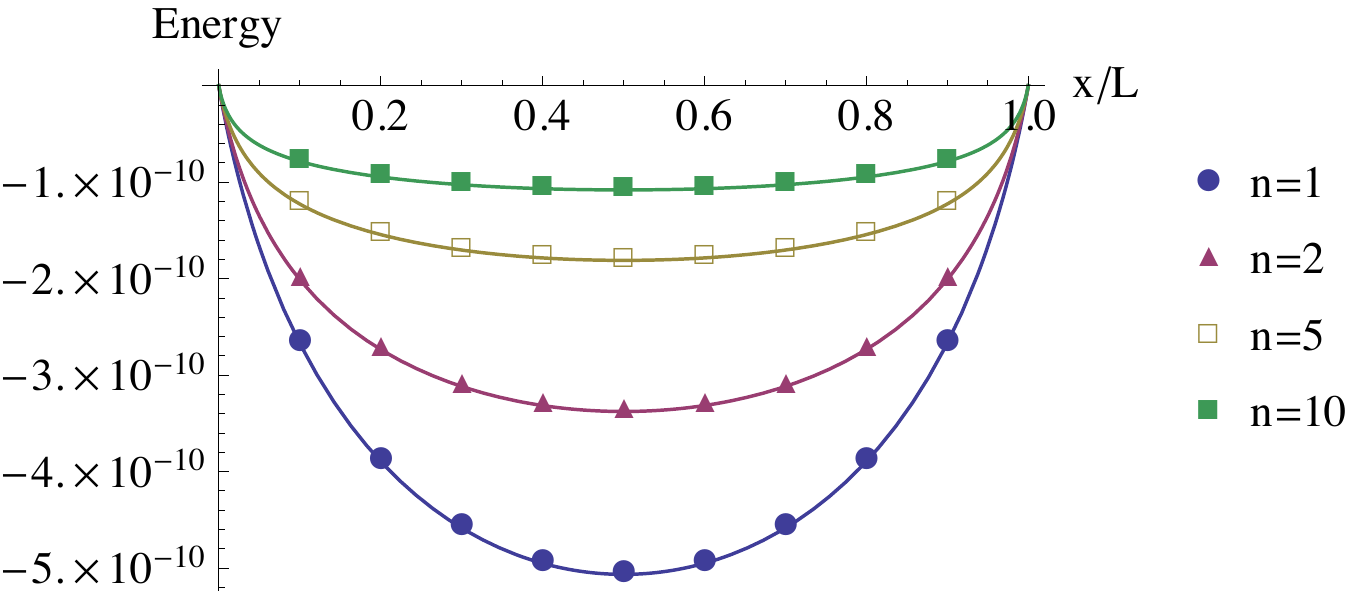}
\caption  {Energy (in units of $1/L$) of the new ground state of the cavity-atom system, to second order, as a function of the position of the atom in the cavity. As seen by the form of the curve, the plates exert a repulsive force in the detector towards the centre. The curves correspond to different sizes of the atom energy gap. In each curve the energy gap of the atom is resonant with one of the modes in the cavity. The smaller the energy gap $\Omega$ is, the lower the energy of the new ground state will be. We chose the parameters $L=1$ and $\lambda=10^{-4}$ (all energies in units of $1/L$).} \label{fig:En}
\end{figure}

The functional shape of the energy for various atomic gaps is plotted in Fig.\ref{fig:En}. We see that the atom experiences a repulsive Casimir-type force, away from the cavity walls, in the direction of the middle of the cavity. In comparison, the calculation for an atom close to a plate, taking into account both electric and magnetic contributions with their respective boundary conditions, yielded a net attractive force \cite{CasimirPolder}.

Let us also consider the interaction energy's dependence on $L$, the size of the cavity. Namely, in addition to the original attractive Casimir force  \cite{Casimir}  between two conducting plates, there will be an effect due to the presence of an atom between the plates. We can calculate that additional force due to the presence of the atom by differentiating the total energy with respect to the parameter $L$.

\begin{figure*}

\includegraphics[width=0.42\textwidth]{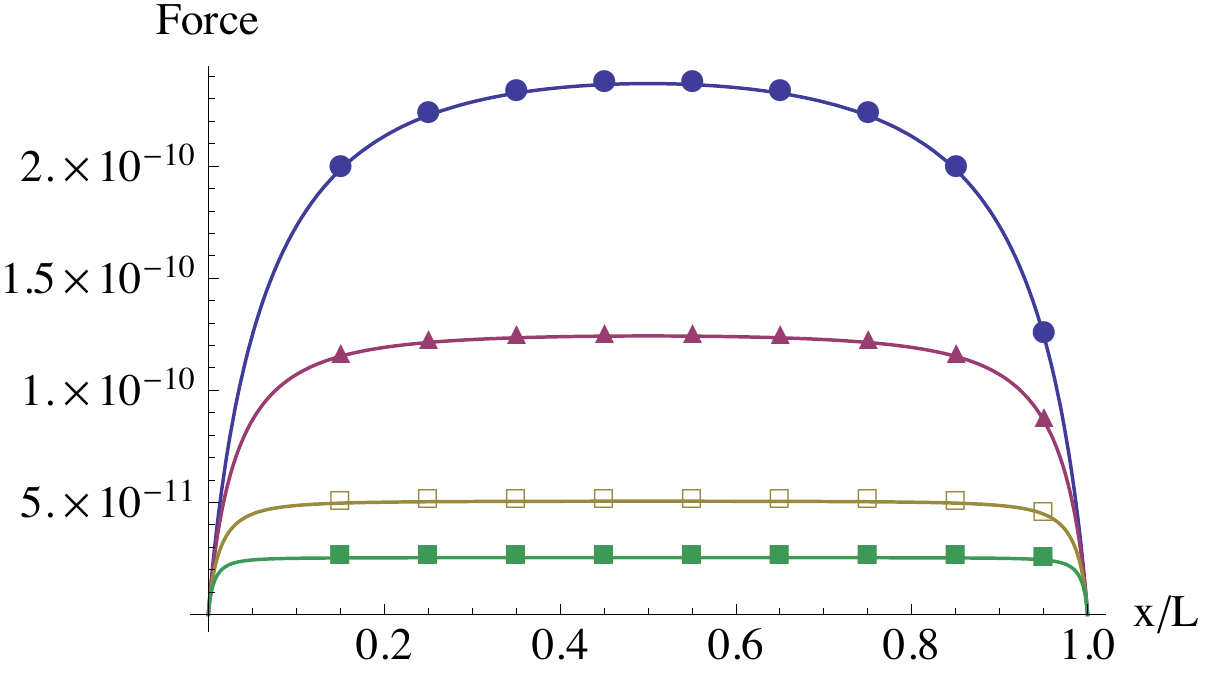} \label{fig:ForxL}
\includegraphics[width=0.48\textwidth]{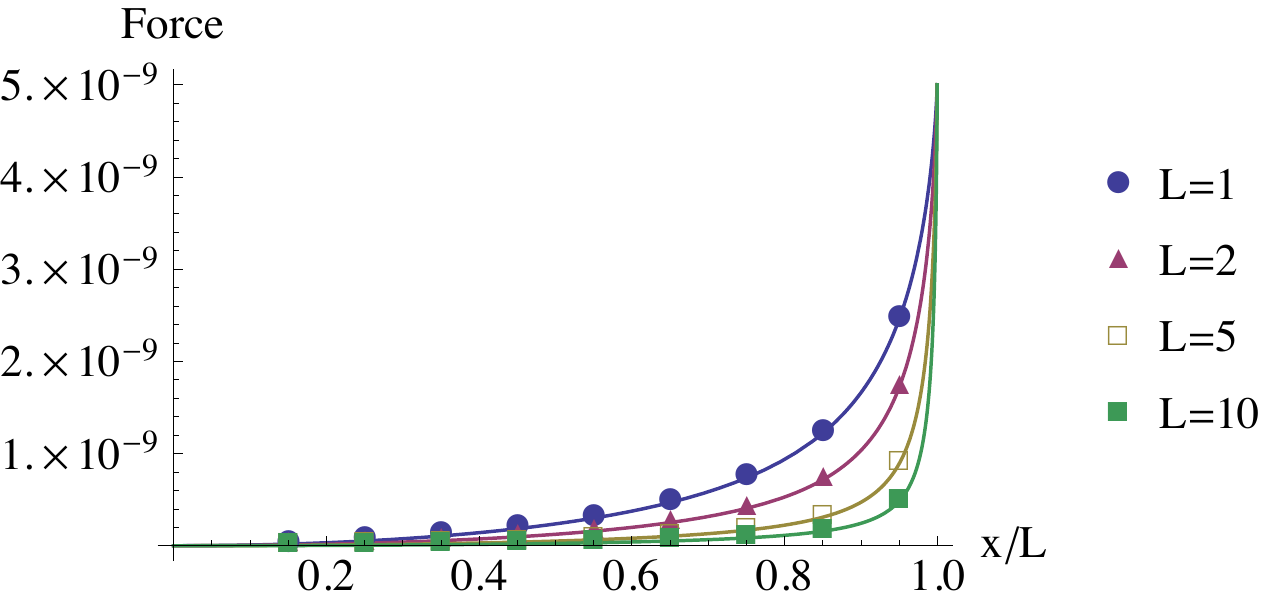} \label{fig:Forx}
\caption  {The left graph shows the repulsive force (in units of $1/L^2$) on the plates in the case in which the atom conserves its relative position to both plates when they move. The right one shows the repulsive force (in the same units) on the plate at $x=L$ when we fix the distance between the atom and the boundary at $x=0$ as the further one is moved. This differentiation between the plates is the cause of the lack of symmetry in the second graph. At $x/L=1$, all the curves converge to the same value. Both forces are plotted as a function of the position of the atom within the reflective cavity. We note how in the two the force decreases with the size of the cavity. For both these figures we choose $\Omega=2\pi$ and $\lambda=10^{-4}$ (all energies in units of $1/L$). } \label{fig:ForxxL}
\end{figure*}

However, there are two very different ways in which we can differentiate it, which are $i)$ fixing $x/L$ constant or $ii)$ fixing $x$ constant. The first case corresponds to a situation in which the relative position of the atom is kept constant within the cavity as the length varies infinitesimally. It is a case that will be natural to consider later, when we will introduce a large number of atoms into the cavity so as to model a dielectric medium. The second case yields the force occurring when only one of the two plates is allowed to move infinitesimally, while the atom and the other plate remain at fixed positions with respect to each other.  
The expression of the force in each case is, respectively,
\begin{align} \label{eq:FoxL}
F_{x/L}=\sum_{j=1}^\infty \frac{\lambda^2 \sin^2{\frac{\pi j x_\text{d}}{L}}}{(\pi j+L \Omega)^2} \\ \label{eq:Fox}
F_x=F_{x/L}-\sum_{j=1}^\infty \frac{\lambda^2 x_\text{d} \sin{\frac{2 \pi j x_\text{d}}{L}}}{(\pi j+L \Omega)L}.
\end{align}
When setting $x/L$ constant we get an expression for the force that is always positive, meaning that it will be a repulsive force opposing the usual attractive Casimir forces between the plates, as well as symmetric over the cavity as we can see in Fig.\ref{fig:ForxxL}. For the force in the second situation we break the symmetry between the plates when differentiating with respect to $L$. However, as shown again in Fig.\ref{fig:ForxxL}, it is still always repulsive, and stronger the closer the atom is to the moving boundary. 

\subsection{Contribution of the diamagnetic term} \label{sec:phisqrd}
The same analysis may be repeated this time for the interaction Hamiltonian (\ref{eq:Hi2}) which includes the diamagnetic term.

Now, the leading order correction (in $\lambda$) to the ground state energy has two contributions. There is one contribution in second order in perturbation theory which is similar to (\ref{eq:secen}) coming from the $\lambda m_{\text{d}}\phi$ term. There is also a contribution which is first-order in perturbation theory coming from the $\lambda^2 \phi^2$ term. We will call these two contributions $ E^{(2)}_\text{UdW}$ and $ E^{(1)}_{\phi^2}$ respectively. Both contributions are  $\mathcal{O}(\lambda^2)$, so they contribute equally to the energy shift, which in this case takes the form:
\[\delta E=  E^{(2)}_\text{UdW}+ E^{(1)}_{\phi^2}+\mathcal{O}(\lambda^4)\] 

As we discussed earlier, we can now no longer use the approximation that the atom couples to the field only at one point because then the $ E^{(2)}_{\phi^2}$ of the self-interaction energy would be divergent. This was observed already in the original work of Casimir and Polder \cite{CasimirPolder}.

Hence, we need to consider that the atom has a physical extension and model a spatial profile for it.  Here we will choose as a spatial profile the square of the wave function of the ground state of atoms such as hydrogen, in the same fashion as in \cite{Wavepackets}. Hence, if $a_0$ is some characteristic length scale (the Bohr radius or a characteristic radius of a spherically symmetric atomic species) then:
\begin{equation} \label{eq:Profile}
\Psi(x)=\frac{e^{-x/a_0}}{a_0}
\end{equation}
 It can be easily seen (and as is also explained in detail, e.g., in  \cite{Wavepackets}), that this spatial profile affects the Hamiltonian by introducing a weighted interaction between the atom and the field modes, with the weight equal to the Fourier transform of the spatial profile (\ref{eq:Profile}). Here, this is
\[f(k,a_0)=\frac{2}{(a_0 k)^2+1}.\]
Since we are summing over a discrete number of modes, the weight per mode is then:
\begin{equation}\label{weeeeitea}
f_{j}(L,a_0)=\frac{2}{(a_0\frac{ \pi  j}{L})^2+1}
\end{equation}
In \cite{CasimirPolder}, the integral over the momentum is regularized by the introduction of the factor $e^{-\gamma k}$ with the limit $\gamma\rightarrow0$ taken eventually. While this limit constitutes the point-like approximation for the atom, we here take a somewhat more realistic approach by instead regularizing the momentum integrals through a finite Lorentzian atomic spatial profile, which is typical for orbitals of zero angular momentum. In our toy model, for simplicity, we will choose the same profile for the ground and excited states. As we will discuss below, this model, with the appropriate choice of $\alpha$, also describes atomic transitions of the type $1s\rightarrow 2s$.

 Additionally, in our simplified model we are considering that both, the diamagnetic and paramagnetic terms are coupled to the charge density (spatial smearing of the atomic orbitals). However, if we assume that our $m_\text{d}$ operator comes from a restriction to two levels of the momentum operator, one has to be careful with the fact that this term does not couple the field  to the charge density but to the `current density'. Indeed, in our simple model we consider that
\[m_\text{d}=\ket{g}\bra{e}+\ket{e}\bra{g},\]
which means that we are giving a weight 1 to the non-diagonal matrix elements of $p$. Assuming that the right energy scale is given by the choice of $\lambda$, the actual form of our operator $m_\text{d}$ should instead be
\begin{equation}
m_\text{d}\propto \bra{e}p\ket{g}\proj{e}{g}+\bra{g}p \ket{e}\proj{g}{e},
\end{equation}
where in the position representation, the radial momentum is $p=\ii\left(\partialfrac{}{r}+\frac{1}{r}\right)$. Thus, in the case of an electronic transition between two levels of an atom which have rotationally invariant symmetry, these matrix elements will be given by the integral
\[4\pi \ii \int_0^\infty \psi_{1s}(r)\left(\partialfrac{\psi_{2s}(r)}{r}+\frac{\psi_{2s}(r)}{r} \right) \text{d} r,\]
where,  in the case of a $1s\rightarrow 2s$ transition, the two wavefunctions are:
 \[\psi_{1s}=\sqrt{\frac{1}{\pi a_0^3}} e^{-\frac{r}{a_0}},\qquad  \psi_{2s}=\sqrt{\frac{1}{8 \pi a_0^3}}\left(1-\frac{r}{2 a_0}\right) e^{-\frac{r}{2 a_0}}.\]

The difference between our toy model and the electromagnetic coupling between different  atomic electronic levels is a constant factor (dependent on the length-scale of the atom $a_0$) that can, in principle, be reabsorbed in the definition of $\lambda$. However, this is a rather more subtle matter, since the same $\lambda$ also appears in the diamagnetic term, where, in the position representation, no derivative operators are involved (roughly speaking, this term only involves the charge density). Hence, for our toy model to be representative of a realistic atomic transition scenario, we should be able to compensate for that factor by choosing a suitable value of $\alpha$. In this fashion, it is the scale $a_0$ and the orbitals that represent the ground and excited states, which determines, to some degree, what adjustments need to be done to the value of $\alpha$ to be able to reproduce qualitative features of atomic systems through the UdW model.

For example, if the ground state is an hydrogenic $1s$ level and the excited state is the $2s$ level, then the relative weight $\alpha$ has to correct for a factor 
\begin{equation}
\left| \bra{e}p\ket{g} \right| \!= 4\pi\!\! \int_0^\infty\!\!\!\! \psi_{1s}(r)\left(\partialfrac{\psi_{2s}(r)}{r}+\frac{\psi_{2s}(r)}{r} \right) \text{d} r =\frac{ 4 \sqrt{2}}{27 a_0},
\end{equation}
which, for $a_0=10^{-2}$, is $\sim20$. This can be compensated for in our toy model by a suitable
choice of $\alpha$ in Eq. \eqref{eq:Hi2} (which would be $\alpha \sim \frac{1}{20^2}\alpha'$ where $\alpha'$ will be the value of this constant chosen in our model). 

There is no accurate value for $\alpha$, as our UdW model is not meant to accurately describe the full 3-dimensional electromagnetic system. Instead, by using a simple but fully manageable UdW model in one dimension our aim is to explore the range of possible qualitative contributions of the paramagnetic and diamagnetic terms to the Casimir-Polder force.   
We explore the Casimir Polder forces for a range of values for $\alpha$ in Appendix \ref{appfig}. Indeed, as shown in Fig. \ref{fig:Alpha} we find that a small enough value of $\alpha$ leads to a qualitatively new behavior in that it changes the sign of the forces. Combined with our discussion of what influences $\alpha$ in the UdW model, this illustrates that the direction of the Casimir-Polder force can depend on the geometry of the atoms involved.

As mentioned above, in order for our UdW model to possess a realistic relative size of the paramagnetic and diamagnetic terms in atomic systems, a natural choice is to set $\alpha=1$. This choice can be readily shown to effectively provide factors that reasonably model the full electromagnetic case in 3D. To see this, one can compare our expressions \eqref{eq15} and \eqref{eq:secen2} with the analogous terms obtained in the 3D electromagnetic case with the proper summation rules in  \cite{CasimirPolder}. For simplicity, throughout the rest of the calculations we will set $\alpha$ to either $0$ (when we want to study only the role of the paramagnetic term) or $1$ (when the diamagnetic term dominates as in the full electromagnetic term in 3D \cite{CasimirPolder}). 



\begin{figure}
\includegraphics[width=0.48\textwidth]{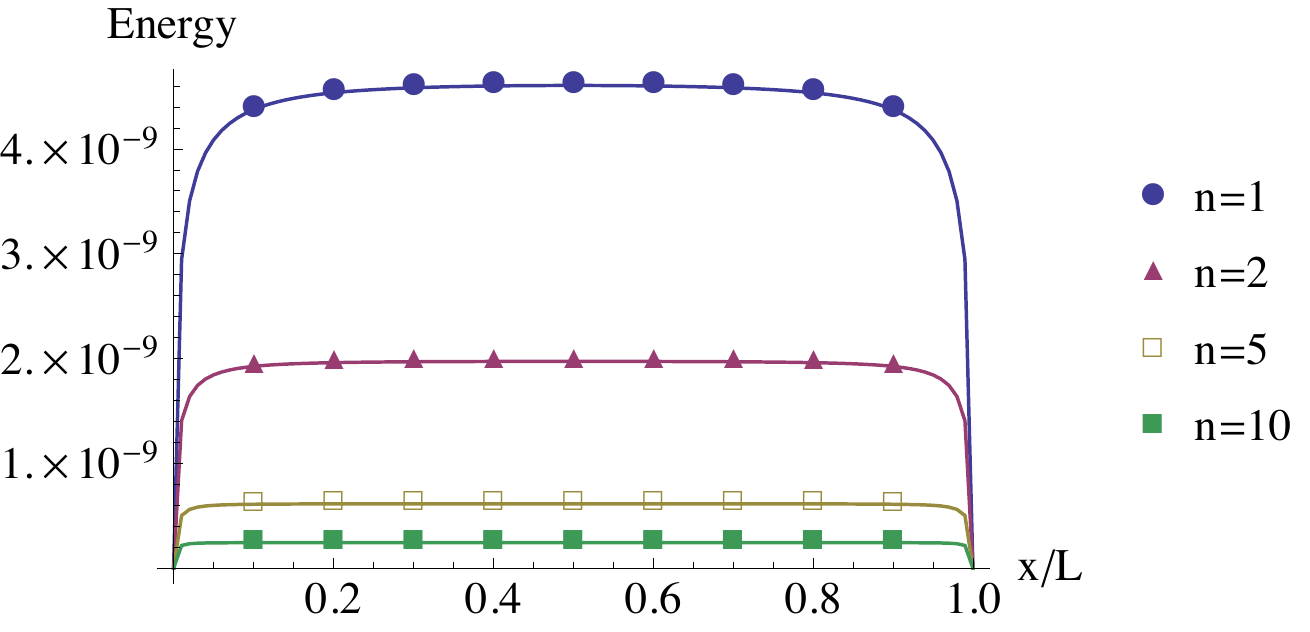}
\caption  {Energy (in units of $1/L$) of the new ground state, to second order, in the case in which we include the self-interaction of the field with a spatial profile. We can see that here there is an attractive force towards the plates, as we find in the classic treatment of the Casimir-Polder force. The energy gap was chosen to be resonant with different field modes, and we choose the parameters $L=1$, $\lambda=10^{-4}$ and $a_0=10^{-3}$ in natural units (with L setting the units of length and all energies in units of $1/L$). Again we note how the coupling energy decreases with increasing energy gap.} \label{fig:EnSI}
\end{figure}

Modifying the interaction to take into account the atomic spatial profile \eqref{eq:Profile}, hence changes the Hamiltonian \eqref{eq:Hi2} to:
\begin{align}\label{eq:Hi2SI}
H_{I}'& =  \lambda\,  m_\text{d} \sum_{j=1}^{\infty} f_{j}(L,a_0) (a_{j}^{\dagger}+a_{j} )\frac{\sin{k_{j} x_\text{d}}}{\sqrt{\omega_{j}L}} \nonumber \\
&+\frac{\lambda^2}{\Omega} \Big[\sum_{j=1}^{\infty} f_{j}(L,a_0) (a_{j}^{\dagger}+a_{j} )\frac{\sin{k_{j} x_\text{d}}}{\sqrt{\omega_{j}L}}\Big]^2
\end{align}
And we then have:
\begin{align}
\label{eq15}
 E^{(2)}_\text{UdW}&=-\lambda^2\sum_{j=1}^{\infty}\frac{  [f_{j}(L,a_0)]^2 \sin^2k_j x_\text{d}}{(\omega_j+\Omega)(\omega_j L)}
\end{align} 
\begin{align}
 E^{(1)}_{\phi^2}&= \lambda^2 \sum_{j=1}^{\infty}\frac{[f_{j}(L,a_0)]^2\sin^2{k_{j} x_\text{d}}}{\Omega \, \omega_{j} L} \nonumber
\end{align}
Adding the two we obtain for the energy to second order:
\begin{align} \label{eq:secen2}
\delta E= E^{(2)}_\text{UdW}+ E^{(1)}_{\phi^2}=\lambda^2 \sum_{j=1}^\infty \frac{[f_{j}(L,a_0)]^2 \sin^2{k_{j} x_\text{d}}}{ \Omega L (\Omega+\omega_{j})}
\end{align}
Again this series can be analytically summed into a closed expression. The analytical expression of this series is given in Appendix \ref{EnergySI}. It turns out that, as we can see in Fig.\ref{fig:EnSI}, the shape of the energy curve is the opposite in sign to what we can see in Fig.\ref{fig:En}, telling us that there will be an attractive force that will pull the atoms towards the plates, in a similar way to \cite{CasimirPolder}.  This shows that  the nature of the Casimir-Polder forces is highly affected when we incorporate the $\phi^2$ term to our model, and without it the forces would be repulsive instead of attractive.

Now again, this energy depends on the length of the cavity, which means that a force will appear onto the walls of the cavity due to the presence of atoms, in the same way it did under variations of the position of atoms in the cavity. In this scenario we again also consider the differentiation under the two different dynamical constraints. When $x_\text{d}/L$  is fixed we have a force given by
\begin{align} \label{eq:ForcePSI}
&F_{x/L}=-\frac{d E^{(2)}}{d L}= \sum_{j=1}^\infty b_{j}^\text{d}
\end{align}
Here, we defined:
\begin{align}
&b_j^\text{d}=\frac{4 \lambda ^2 L^3 \left(L^3 \Omega -\pi ^2 a_0^2 j^2 (4 \pi  j+3 L \Omega )\right) \sin ^2\left(\frac{\pi  j x_\text{d}}{L}\right)}{\Omega  \left(\pi ^2 a_0^2 j^2+L^2\right)^3 (\pi  j+L \Omega )^2}.
\end{align}

When we fix $x_\text{d}$ instead we have
\begin{align}  \label{eq:ForcePSI2}
F_x=F_{x/L}+\sum_{j=1}^\infty\frac{\lambda^2 f_{j}^2(L,a_0) \pi j x_\text{d} \sin{(\frac{2 \pi j x_\text{d}}{L})}}{\Omega L^2 (\pi j+L \Omega)}
\end{align}

\begin{figure*}
{\label{fig:ForSIN1}
\includegraphics[width=0.48\textwidth]{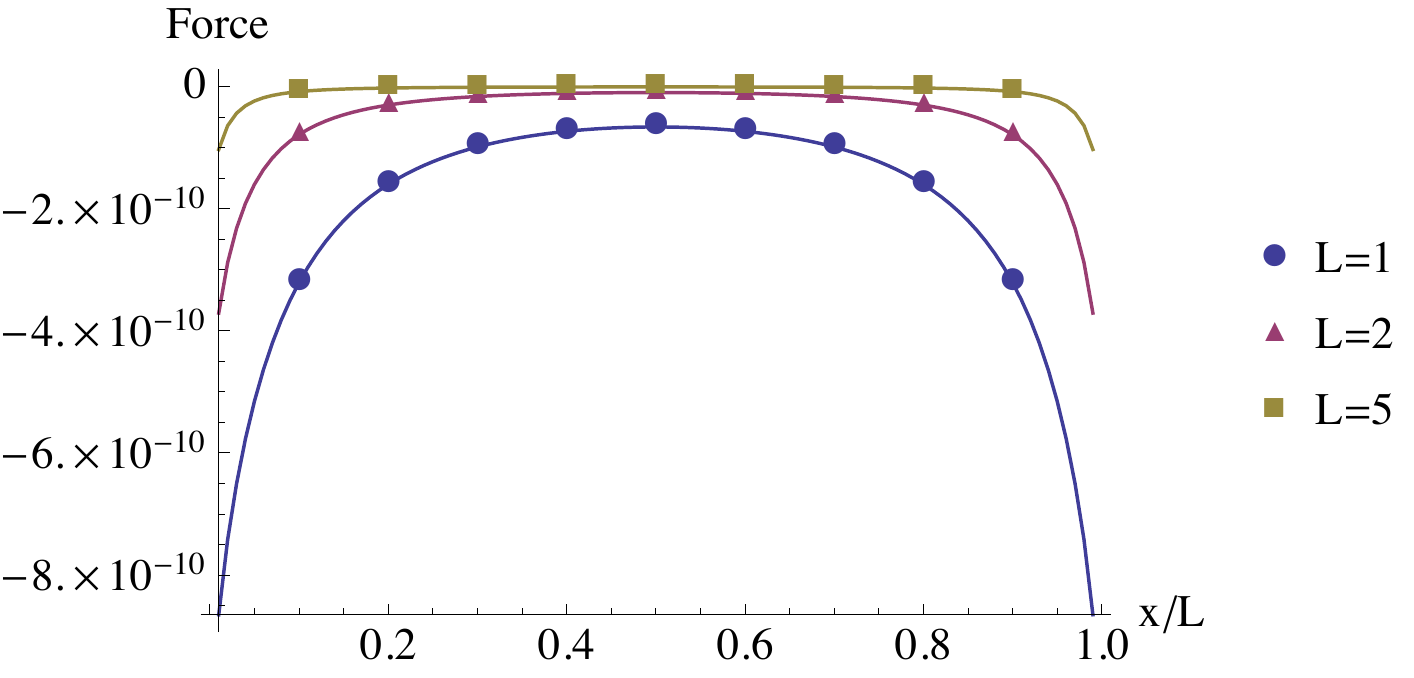}}
{\label{fig:ForSIN2}
\includegraphics[width=0.48\textwidth]{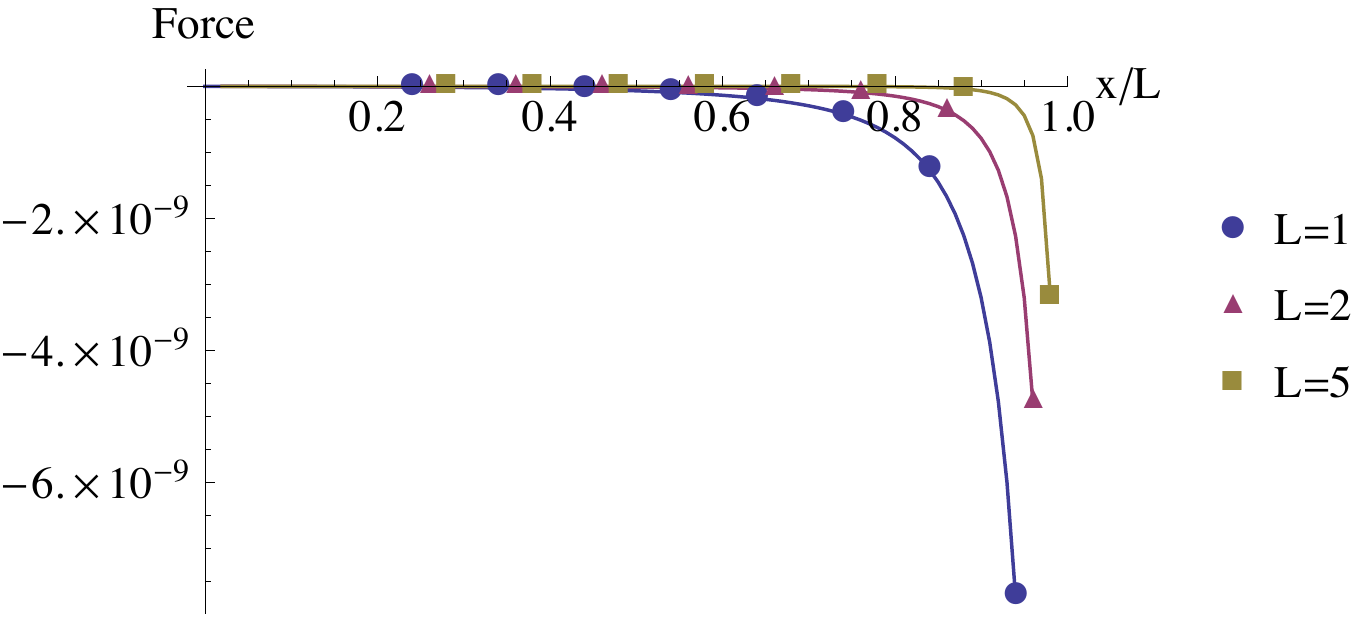}}
\caption{Plots of the force (in units of $1/L^2$) of a single atom on the walls including the $\phi^2$ term. The left plot shows the force when the atom is fixed at $x_\text{d}/L$ as given by (\ref{eq:ForcePSI}). The right one is the force on plate at $x=L$ from the atom fixed at $x_\text{d}$, as given by (\ref{eq:ForcePSI2}). Now, this force, being always negative, is attractive between the plates in both cases. The lines in the second plot converge to a highly negative value, different for each one of them, when approaching $x_\text{d}=L$. For both figures we choose the parameters $L=1$, a Bohr radius of $a_0=10^{-3}$,  and $\lambda=10^{-4}$ (with $L$ setting the units of length and all energies in units of $1/L$).} \label{fig:ForSIN}
\end{figure*}

The magnitude of these forces is shown in Fig.\ref{fig:ForSIN}, where we can see that the contribution is here always attractive, as opposed to the case where the interaction did not have a $\phi^2$ term.

\subsection{An atom in a Neumann-type cavity}
It is interesting now to analyze how the boundary conditions imposed on the cavity modify the nature of the Casimir-Polder forces, also because in the case of full QED, the electric and magnetic fields tend to obey differing boundary conditions. If in our scenario with scalar fields we consider Neumann conditions instead of reflective boundaries, the nature of the field modes is still quite similar, as the spatial behavior of the modes changes from a sine to a cosine. One may expect, therefore, little to no change in Casimir-type effects. For instance, it has been proven that varying the boundary conditions has no effect in phenomena like the Unruh effect \cite{Wilson}. However, we will see that for the Casimir-Polder effect, the boundary conditions critically change the sign of the force, from attractive to repulsive.

To this end, we repeat the calculation from the previous subsection, now with the solutions to the Klein-Gordon equation having the boundary condition to $\frac{d u_{j}(0,t)}{d x}=\frac{d u_{j}(L,t)}{d x}=0$. As mentioned above, the way this changes the interaction Hamiltonians is simply
\begin{equation}\label{eq:HiVN}
H_{I} =  \lambda\,  m_\text{d} \sum_{j=1}^{\infty} (a_{j}^{\dagger}+a_{j} )\frac{\cos{k_{j} x_\text{d}}}{\sqrt{\omega_{j}L}}
\end{equation}
for the standard Unruh-DeWitt model and
\begin{align}\label{eq:Hi2VN}
H_{I}' =  \lambda\,  m_\text{d} \sum_{j=1}^{\infty}  f_{j}(L,a_0) (a_{j}^{\dagger}+a_{j} )\frac{\cos{k_{j} x_\text{d}}}{\sqrt{\omega_{j}L}} \nonumber \\+\frac{\lambda^2}{\Omega} \Big[\sum_{j=1}^{\infty}  f_{j}(L,a_0)(a_{j}^{\dagger}+a_{j} )\frac{\cos{k_{j} x_\text{d}}}{\sqrt{\omega_{j}L}}\Big]^2
\end{align}
for the Hamiltonian with the diamagnetic term.

The expression for the energy corresponding to (\ref{eq:HiVN}) is
\begin{align} \label{eq:secenVN}
\nonumber E^{(2)}=&\sum_{l=1}^{\infty}\frac{-1}{\omega_{l}+\Omega}\Big|\bra{e,k_{l}}\lambda\, m_d \sum_{j=1}^{\infty} (a_{j}^{\dagger}+a_{j} )\frac{\cos{k_{j} x_\text{d}}}{\sqrt{\omega_{j}L}}
\ket{g,0}\Big|^{2} \\
=&\sum_{j=1}^{\infty}\frac{-\lambda^{2} \cos^2\frac{\pi jx_{d}}{L}}{(\pi j/L+\Omega)(\pi j)}
\end{align} 
for which there is an analytical expression given by
\begin{align}
&E^{(2)}=-\frac{\lambda ^2}{4 \pi  \Omega } \left[2 {\cal{H}}\left({\frac{L \Omega }{\pi }}\right)-e^{\frac{2 i \pi  x_\text{d}}{L}} \Phi \left[e^{\frac{2 i \pi  x_\text{d}}{L}},1,\frac{L \Omega }{\pi }+1\right] \right. \nonumber \\ &\!\! \left. +e^{-\frac{2 i \pi  x_\text{d}}{L}} \Phi \left[e^{-\frac{2 i \pi  x_\text{d}}{L}},1,\frac{L \Omega }{\pi }+1\right]\!-\!\log \left(2\!-\!2\cos{\frac{2 \pi  x_\text{d}}{L}}\right)\right]
\end{align}
The energy for the Hamiltonian in (\ref{eq:Hi2VN}), when considering the self-interaction of the field is, in analogy with (\ref{eq:secen2})
\begin{align} \label{eq:RegSelfEVN}
\delta E=\lambda^2 \sum_{j=1}^\infty \frac{f_{j}^2(L,a_0) \cos^2{k_{j} x_\text{d}}}{ \Omega L (\Omega+\omega_{j})}
\end{align}
An analytic expression for this energy is given in Appendix \ref{EnergySI}.
\begin{figure*}
{\label{fig:EnVN1}
\includegraphics[width=0.48\textwidth]{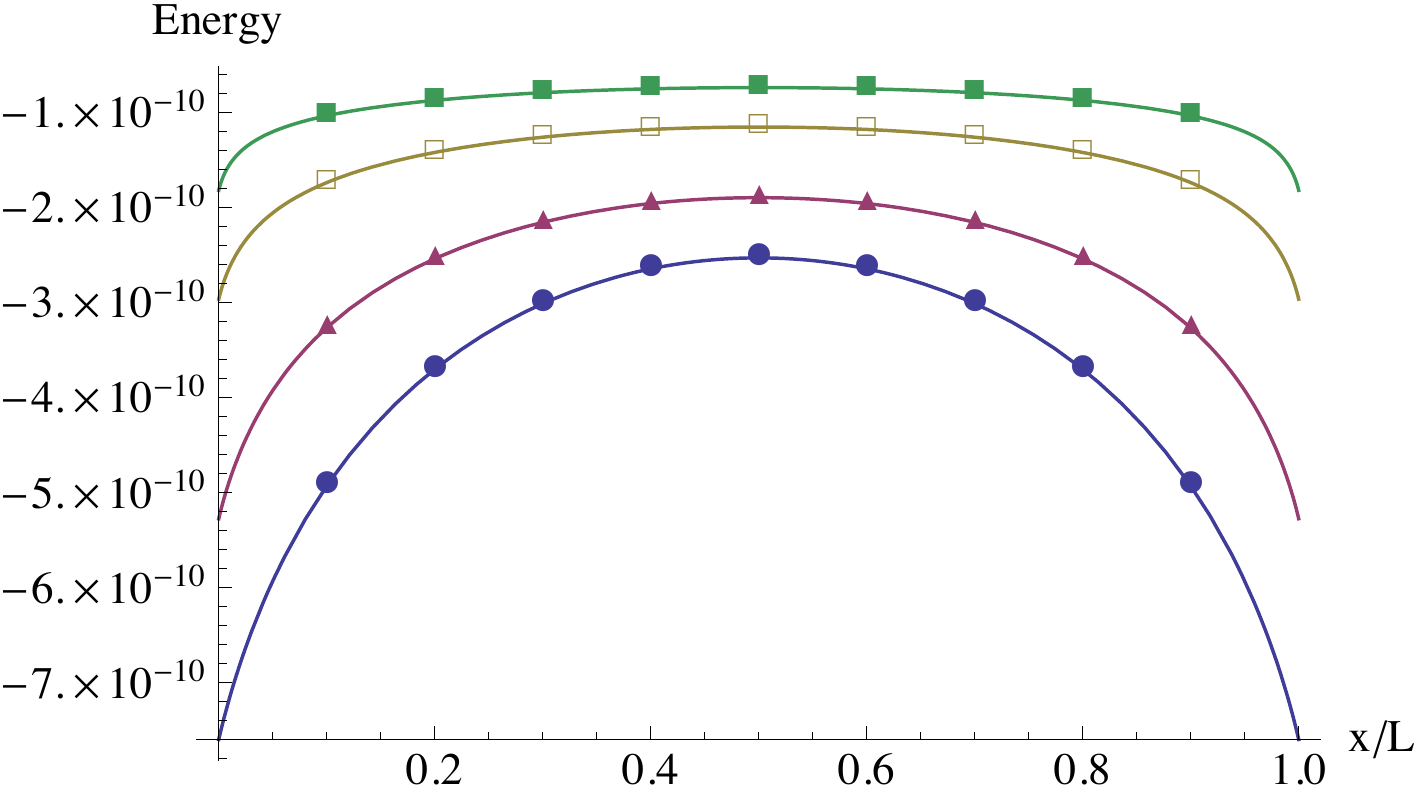}}
{\label{fig:EnVN2}
\includegraphics[width=0.48\textwidth]{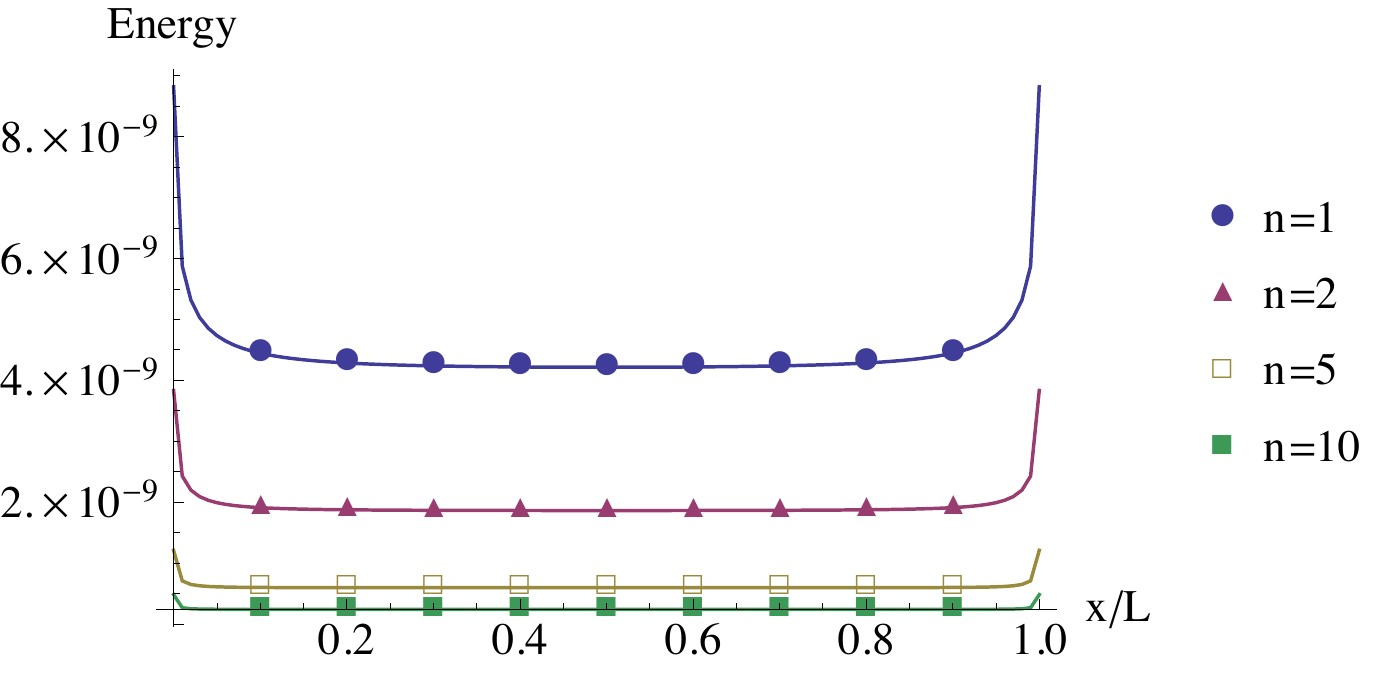}}
\caption{The left plot shows the energy of the ground state (in units of $1/L$) to second order in the Neumann cavity  without the self-interaction energy of the field, to second order, as given by (\ref{eq:secenVN}). Here we have, as opposed to the Dirichlet case, an attractive force on the atom from the plates. The right one shows what happens when we the include the self-interaction energy, from (\ref{eq:RegSelfEVN}), where it can be seen how the sign of the Casimir-Polder forces is inverted. For both figures we choose the parameters $L=1$ and $\lambda=10^{-4}$ in natural units (with L setting the units of length and all energies in units of $1/L$). For the second graph we choose a Bohr radius of $a_0=10^{-3}$, which explains the difference in energy scale between the two. Again the curves correspond to different sizes of the atom energy gap, where $n$ is the field mode to which the atom is coupled. We can see how in general the larger the size of the energy gap, the smaller the coupling with the field. } \label{fig:EnVN}
\end{figure*}

The effect of this change of boundary conditions is to invert the behavior as compared to the one for a reflective cavity. This can be seen in both plots of Fig. \ref{fig:EnVN}, for the interaction Hamiltonians \eqref{eq:HiVN} and \eqref{eq:Hi2VN}, i.e.,  with and without the diamagnetic term. Also, once again,  the attractive forces in the UdW case turn repulsive when including the diamagnetic term. 

\section{Modeling a dielectric medium in a cavity} \label{sec:VarN}

In this section we analyze the Casimir forces between two conducting plates when a dielectric medium is introduced between the plates; in particular, a medium made up of a large number of uniformly distributed atoms. This simple model of a medium filling the cavity can be dealt with by an extension of the results in the previous sections, given that the leading-order part of the total energy of the system due to the atoms will then just be the sum of all the single-atom contributions. For the $N$ atoms, the energy of the Dirichlet cavity is in the two different cases that we have been treating:
\begin{align} \label{eq:Nsecen}
 \delta E_{N}^{\text{UdW}}=\sum_{n=1}^N \sum_{j=1}^{\infty}\frac{-\lambda^{2} \sin^2{k_{j} x_{n}}}{L \omega_{j}(\omega_{j}+\Omega)}
\end{align} 
\begin{align}\label{eq:Nsecen2}
\delta E_{N}^{\phi^2}=\sum_{n=1}^N \sum_{j=1}^\infty \frac{\lambda^2 f_{j}^2(L,a_0) \sin^2{k_{j} x_{n}}}{ L\Omega (\omega_{j}+\Omega)}
\end{align}
Here, $ \delta E_{N}^{\text{UdW}}$ is the ground state energy shift  considering $N$ atoms coupled through the standard Unruh-DeWitt Hamiltonian that yields (\ref{eq:secen});  and $\delta E_{N}^{\phi^2}$ is the shift obtained after including the $\phi^2$ self-interaction term, which yields (\ref{eq:secen2}). The exact magnitude of these two will of course depend on the specific distribution of the $N$ atoms inside the cavity. For our purposes we will assume a uniform distribution of particles as a rough model of a homogeneous dielectric medium. Our aim is to compute the force exerted onto the walls by the presence of that medium and to determine under what circumstances this force may  overcome the originally attractive Casimir force between the two plates. 

Note that if we introduce a number of atoms in the cavity, then both the ground state energy of the system and consequently the forces that the presence of the atoms exert on the cavity walls, will be in principle modified by many-body effects. Indeed, as has been known since the original work of Casimir and Polder, there exists a fourth-order interaction between neutral atoms even in free space \cite{CasimirPolder}.  In general, these corrections will be of subleading order with respect to the individual effect of every atom on the walls. Nevertheless, if the density of atoms is high then these effects can no longer be neglected. We will further discuss this issue below and we will show that the approximation of not considering many-atoms-interaction contributions to the energy does not change the qualitative features of our main result.


\subsection{Force on the cavity walls of $N$ Unruh-DeWitt atoms} \label{sec:FnUdW}
From the energy given by (\ref{eq:Nsecen}), we can now obtain the force on the walls of the cavity induced by the presence of $N$ point-like Unruh-DeWitt atoms. Without the diamagnetic term in the Hamiltonian, the presence of those atoms within the cavity builds up a repulsive force on the plates that, for a sufficient number of them, will overcome the attractive Casimir force that would exist without the atoms.

Our setup is $n$ atoms uniformly distributed over the cavity, at positions $x_{n}=L\frac{n}{N+1}$. In the same fashion as at the end of section (\ref{sec:OneDUdW}), we have two different ways of differentiating the energy, which are $i)$ fixing the relative position of every atom with respect to the cavity walls $x_n/L$ and $ii)$ fixing the position $x_n$ with respect to the plate at $x=0$.
In which way we choose to differentiate the energy depends on what dynamical constraints we impose on the system. When setting $x_{n}/L$ fixed we are modelling a situation in which the particle medium expands along with the motion of the cavities, filling the space evenly at all times.
On the other hand, fixing $x_{n}$ and differentiating over $L$ means that we have a setup in which the distance between the left plate at $x=0$ and the medium is fixed, and where the force is applied exclusively on the plate at $x=L$. The first scenario is a better model of a dielectric medium that expands filling the free space generated when the cavity expands infinitesimally. This is both because in the second scenario we break the symmetry of the system and because, if we let the plate move more than by an infinitesimal amount, the assumption of a uniform distribution would break down. 

\begin{figure*}
{\label{fig:ForceN1}
\includegraphics[width=0.48\textwidth]{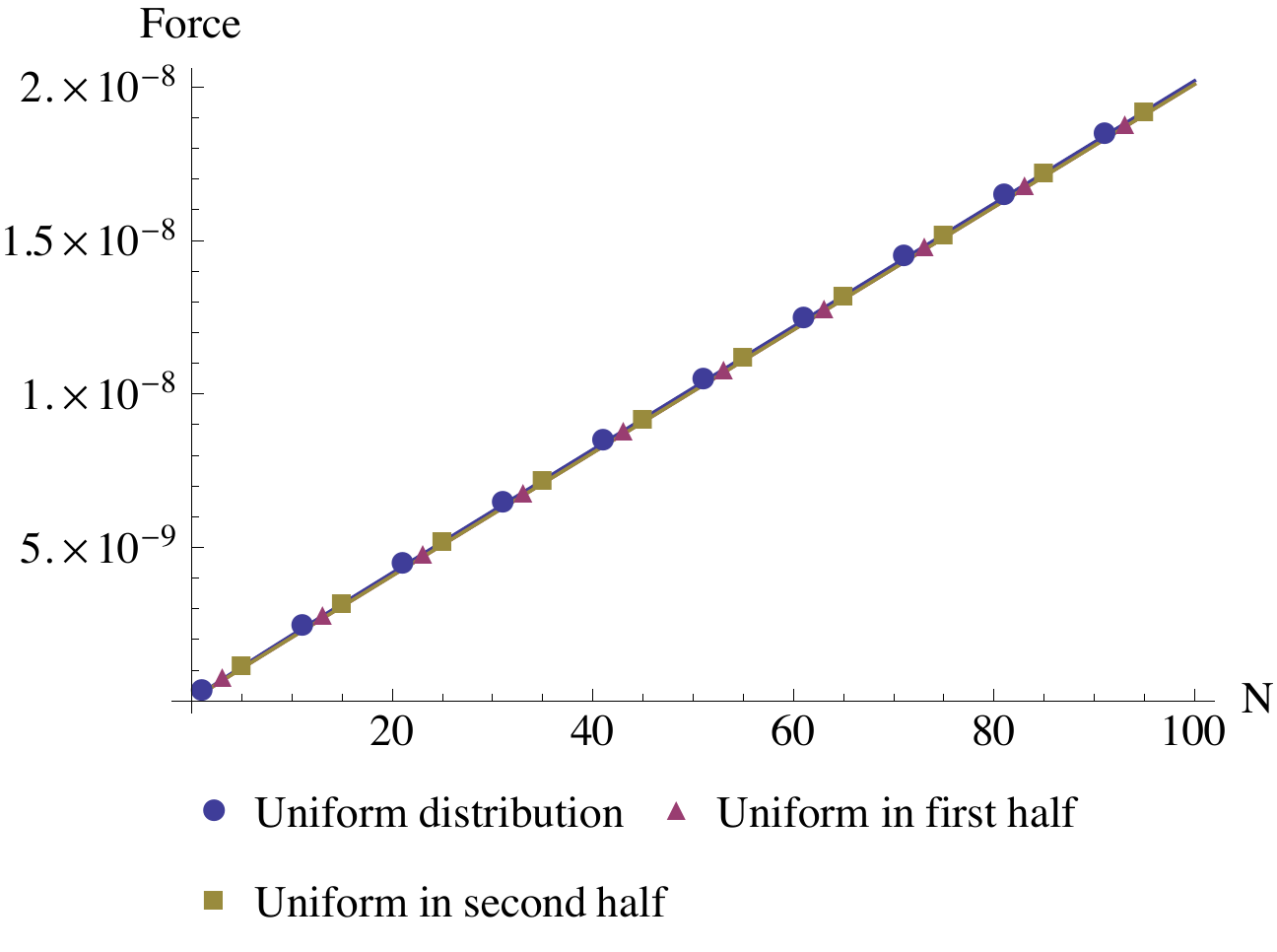}}
{\label{fig:ForceN2}
\includegraphics[width=0.48\textwidth]{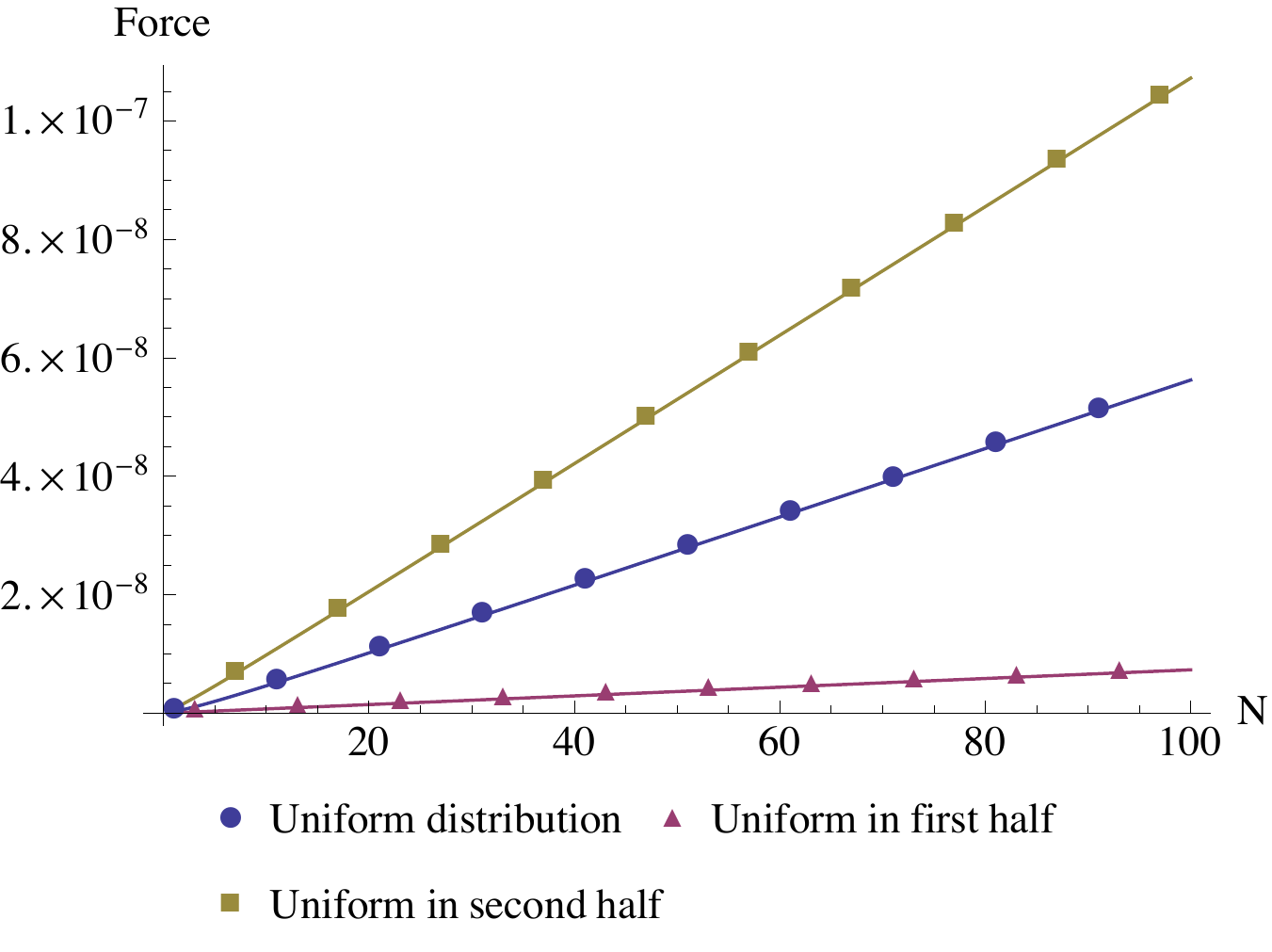}}
\caption{Increase of the repulsive force (in units of $1/L^2$) between the plates of the reflective cavity due to $n$ Unruh-DeWitt atoms when $a)$ the relative position of the atoms $x_{n}/L$ is fixed and $b)$ when we fix the position of them with respect to the wall at $x=0$. In the left one the coincidence of two of the three lines is due to the force being symmetric with respect to $x=L/2$ (the completely homogeneous distribution is slightly different than the half-distributions). In the right one it can be seen how the symmetry between the two plates is broken given the different slopes of the lines. For these we choose $L=1$, $\lambda=10^{-4}$ and $\Omega=2\pi$ (in natural units, with L setting the units of length and all energies in units of $1/L$).} \label{fig:ForceN}
\end{figure*}

 The expression of the force is given by the sum over every single atom in the expressions (\ref{eq:FoxL}) and (\ref{eq:Fox}). 
\begin{align} \label{eq:FxL}
F_{x/L}^N=\sum_n^N\sum_{j=1}^\infty \frac{\lambda^2 \sin^2{\frac{\pi j x_{n}}{L}}}{(\pi j+L \Omega)^2} \\ \label{eq:Fx}
F_x^N=F_{x/L}^N-\sum_n^N \sum_{j=1}^\infty \frac{\lambda^2 x_\text{d} \sin{\frac{2 \pi j x_{n}}{L}}}{(\pi j+L \Omega)L}
\end{align}
In Fig.\ref{fig:ForceN} we show how this force behaves with an increasing number of atoms. In this case the force is repulsive and will tend to separate the cavity walls. For the fixed $x_n/L$, the first case, the forces felt by the walls are relatively insensitive to the distribution of atoms as long as it is uniform. However, in the case where the position with respect to one of the walls is kept constant and only the most distant wall can move infinitesimally, the symmetry is explicitly broken and the force depends on the position of the atoms within the cavity.

\begin{figure}
\includegraphics[width=0.49\textwidth]{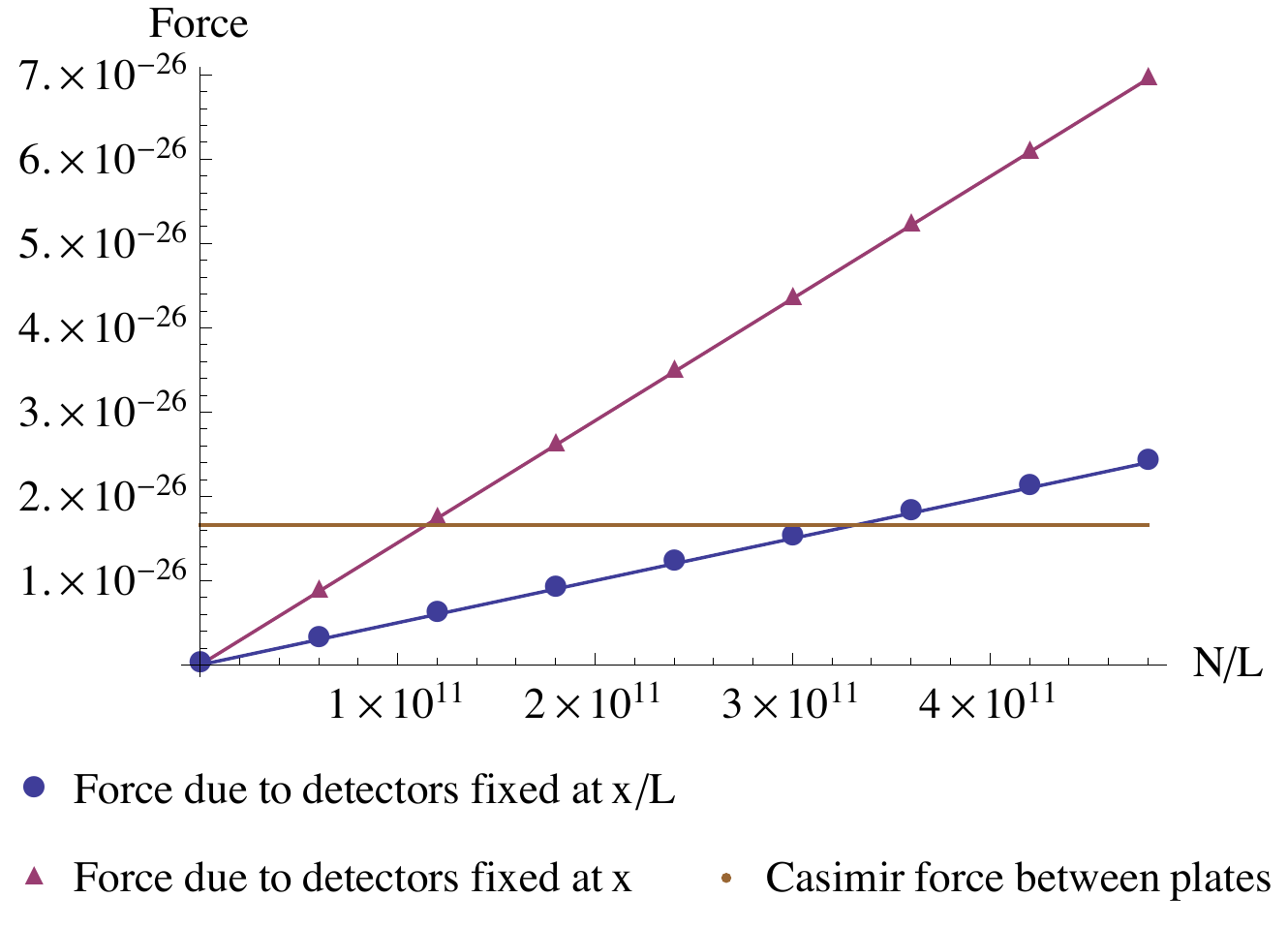} 
\caption{Force (in Newtons) due to a uniformly-distributed row of atoms (at positions $x_{n}/L=\frac{n}{N+1}$), compared with the value of the Casimir force as given by (\ref{eq:CasF}). There are two cases, which are a row of atoms with fixed relative positions $x_{n}/L$, as given by (\ref{eq:FxL}) and the same row of atoms with positions $x_{n}$ fixed with respect to the plate at $x=0$, given by (\ref{eq:Fx}). In both we choose the parameters $L=0.5$ m, $\Omega=\frac{2\pi c}{L}$ resonant with the first field mode and $\lambda=10^{-6}\Omega$. Note that in this graph, unlike in all the others, SI units (Nextons for the force and meters for the length) are used.} \label{fig:CritDen}
\end{figure}

We have seen that in the absence of the diamagnetic term in the Hamiltonian, this contribution to the Casimir force on the walls of the cavity is repulsive. Let us compare this force with the usual attractive Casimir force between the plates, which for a scalar field in 1+1 dimensions is given by  \cite{mukhanov}
\begin{equation}\label{eq:CasF}
F=-\frac{\pi}{24 L^2}
\end{equation}
An immediate question in this setting is then for which linear density of neutral atoms the two opposing forces become equal. This will of course depend on the different energy scales ($\lambda$, $\Omega$ and $L$) and on which of the two differentiation prescriptions is chosen (this is, if we let the two plates move freely or if we keep one fixed and allow the other to move). We show one particular example of this for realistic values in microwave cavities in Fig. \ref{fig:CritDen},  including the two different prescriptions.

We obtain that at a critical density of atoms the plate-plate attractive Casimir forces would be overcome by the Casimir-Polder plate-atoms repulsion. 

\begin{figure}
\includegraphics[width=0.49\textwidth]{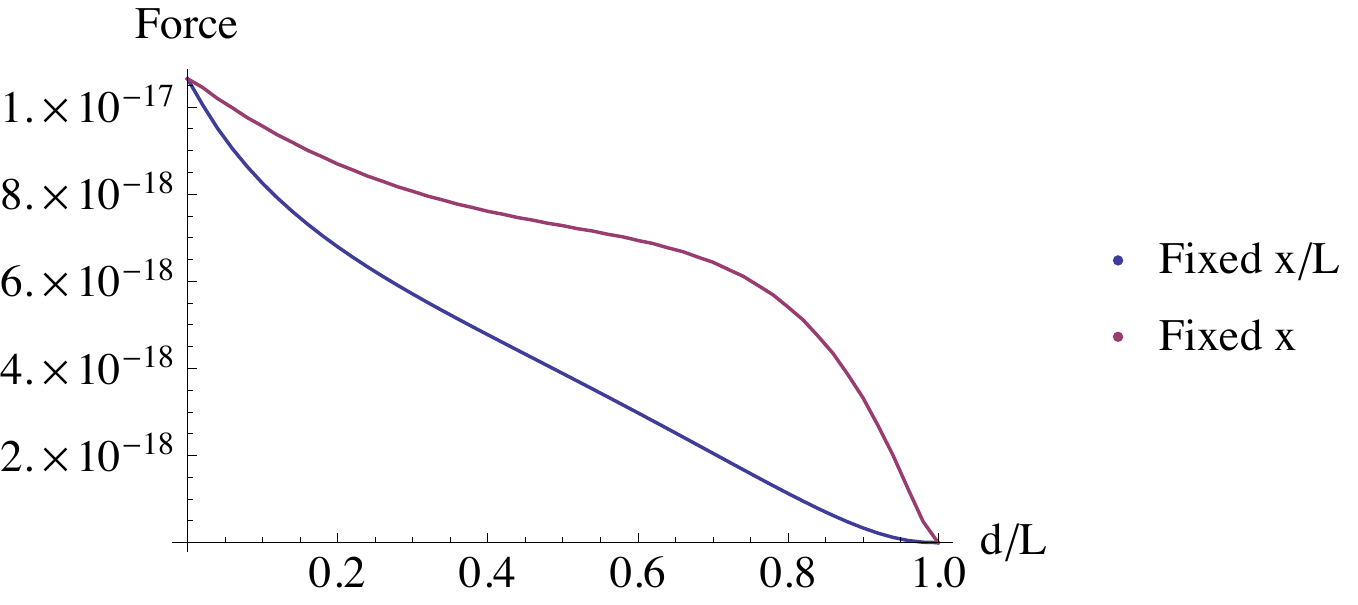} 
\caption{ The force on the plates caused by the interaction of pairs of atoms (in units of $1/L^2$), for the cases of leaving either $x/L$ constant or $x$ constant. It shows that the force is always positive, yielding a repulsive contribution. The two atoms are placed in symmetric positions in the cavity and the horizontal axis represents their distance (i.e., the left of the graph corresponds to both atoms in the middle). The parameters are $\lambda=10^{-4}$, $L=1$ and $\Omega=2\pi$ in natural units, with L setting the units of length and all energies in units of $1/L$.} \label{fig:4thF}
\end{figure}

Note that in considering the expression from  \eqref{eq:CasF} for the force we neglect the impact of the atoms on the eigenstates of the field, which would change it to some extent to become a function of $\lambda$ and the other parameters. This will not, however, change the qualitative behavior. 
\\
Also, as mentioned above, in calculating this we neglect the energy of interaction between the many-body interactions involving many atoms, which is a fourth (and above) order effect in $\lambda$. This is similar to what is known as the (Pairwise summation) PWS approximation in different formalisms \cite{Pablo}. However, this approximation clearly breaks down when $N\sim \lambda^{-2}$. For this reason we additionally calculate the two-body contributions and  show that the contribution of this energy to the force between the plates is also repulsive when we do not consider the diamagnetic term. Thus, there will always be a point at which the attractive Casimir force is canceled out. The expressions for the fourth--order interaction energy between any two atoms and the force when their relative positions are fixed within the cavity (leaving $x/L$ constant) are given in Appendix \ref{4thE}. We show the character of the contribution as a function of the position of the atoms in Fig. (\ref{fig:4thF}), where we can see that this force is always positive, tending to separate the plates.

\subsection{Force on the walls with the self-interaction of the field}
\begin{figure}[h]
{\label{fig:ForceSIN1}
\includegraphics[width=0.48\textwidth]{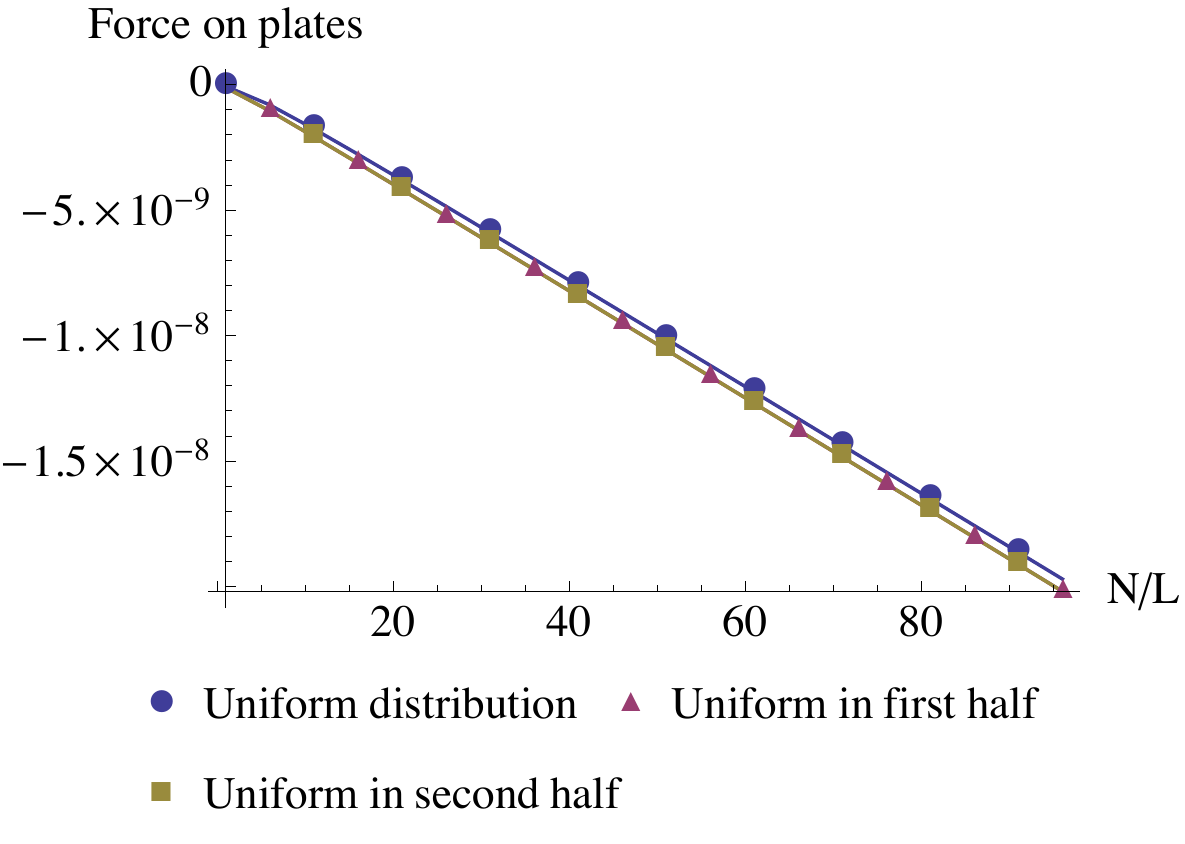} }
{\label{fig:ForceSIN2}
\includegraphics[width=0.48\textwidth]{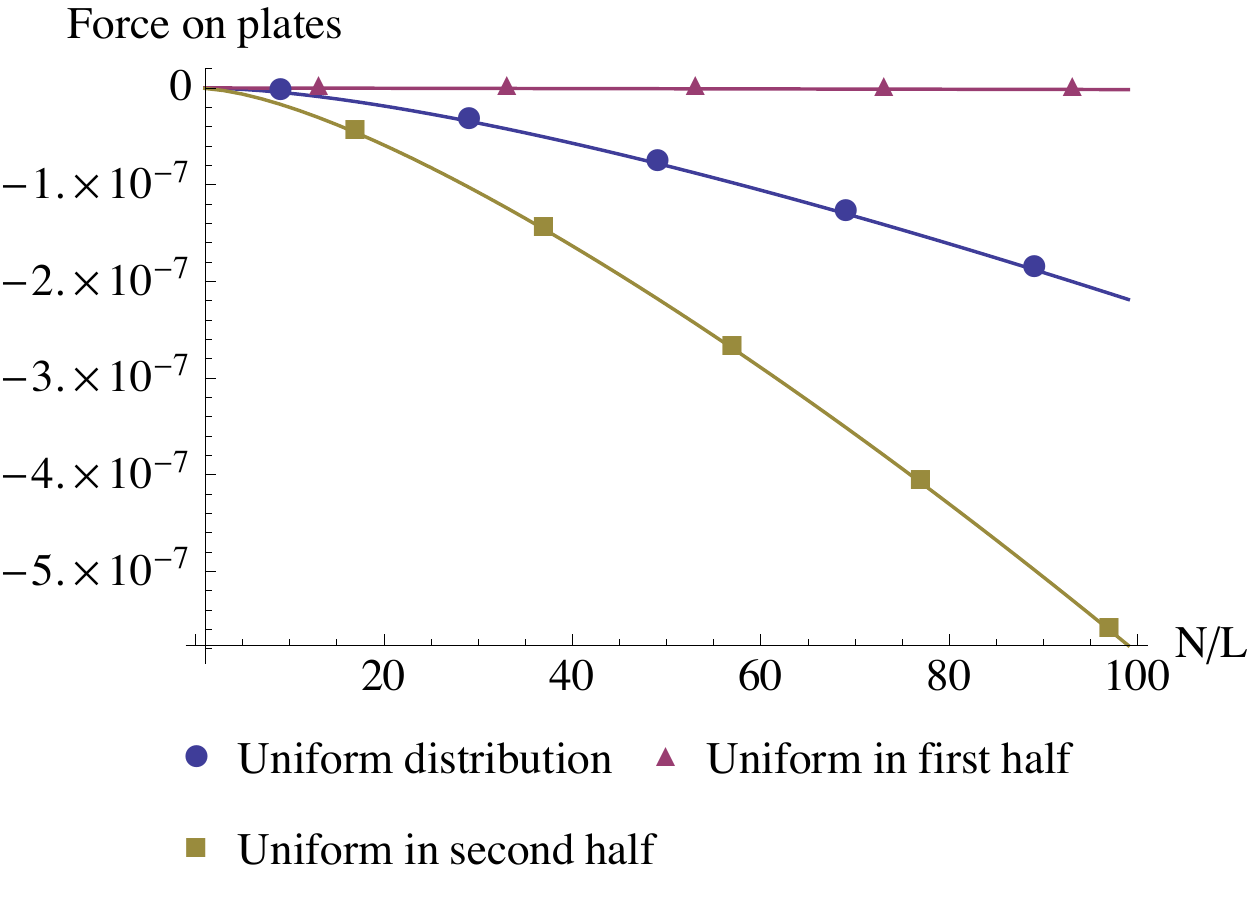}}
\caption{Growth of the attractive force (in units of $1/L^2$) on the plates due to $n$ atoms considering the $\phi^2$ term in $a)$ the situation with the relative position of the atoms $x_{n}/L$ fixed, which is linear (the symmetry of the force with respect to the centre of the cavity is evident by the coincidence of two of the lines) and $b)$ considering the self-interaction energy of the field. The contribution of the asymmetric term is evident in the non-linearity of the blue and yellow lines, distributions in which atoms pile up in the vicinity of the plate at $x=L$. For both graphs we choose $L=1$, $\lambda=10^{-4}$ and $\Omega=2\pi$, and a Bohr radius $a_0=10^{-3}$ (natural units, with energies in units of $1/L$).} \label{fig:ForceSIN}
\end{figure}

The calculations of the previous subsection are repeated this time for the energy as given in (\ref{eq:Nsecen2}) taking into account the field self-interaction term $\phi^2$. The force when fixing the atoms at relative positions $x_{n}/L$ is given by the sum over the $n$ atoms of (\ref{eq:ForcePSI})
\begin{align}
&F_{x/L}^N=-\frac{d \delta E_N}{d L}=\sum_n^N \sum_{j=1}^\infty b_j^n,
\end{align}
where $b_j^n$ is defined above in (\ref{eq:ForcePSI}) for a single atom. When we fix $x_{n}$ constant instead it is the sum of  (\ref{eq:ForcePSI2})
\begin{align}
F_x^N=F_{x/L}^N+\sum_n^N\sum_{j=1}^\infty\frac{\lambda^2 f_{j}^2(L,a_0) \pi j x_{n} \sin{(\frac{2 \pi j x_{n}}{L})}}{\Omega L^2 (\pi j+L \Omega)} \label{eq:ForceNPSI2}
\end{align}
We show the value of these two forces as a function of the number of atoms in Fig.\ref{fig:ForceSIN}. As we showed above in Fig.\ref{fig:ForSIN}  these forces are always attractive, and we hence cannot reproduce a repulsive setup like the one found in the previous section,exactly the opposite to the case where no self-interaction field term is considered

The fact that a medium of atoms is attractive when considering the diamagnetic self-interaction of the field like in the full QED case is consistent with the results of \cite{Lifshitz}. There, it is found that a necessary condition of the repulsive setup is for the medium in the middle to have an electric permitivity with a value in between of those of the materials that make the plates. Having here a cavity of two conducting plates, this is of course not our case.

\section{Conclusions}
We showed that in the light-matter interaction models used to compute Casimir and Casimir-Polder effects it is necessary to carefully account for the diamagnetic term as, unlike in many other instances, this term can even qualitatively change the Casimir-type forces, namely it can turn a repulsive force into an attractive force and vice versa. When the diamagnetic term is present, the atoms are attracted to plates with Dirichlet boundary conditions. But, without the term, the plate-atom forces are repulsive and the atom has a stable equilibrium point at the furthest point between both plates. We also considered the case of an atom with or without diamagnetic coupling term in a cavity with Neumann boundary conditions, in which case the forces are of opposite sign to that of a Dirichlet cavity.

These results suggest a natural and intuitive interpretation. To this end, let us reconsider the interaction of an atom with a finite classical background field. The diamagnetic term leads to a diamagnetic repulsion of the atom from regions of large field strength. This is of course because the diamagnetic term is quadratic in the field, such that the larger the field strength the more energy does the  diamagnetic term add. In contrast, the paramagnetic term is linear in the field and therefore not of definite sign. The atom's degrees of freedom can therefore adjust to the prevailing field to make the energy contribution of the paramagnetic term negative. A paramagnetic term can therefore contribute to a force that attracts the atom towards a region of large field strength. In the cavities that we considered we assumed the vacuum state, i.e, we assumed the absence of any background field so that the atom interacts only with the quantum field's vacuum fluctuations. Does the atom get attracted or repulsed from such vacuum field fluctuations similarly to how it behaves with respect to classical background fields? Our results suggest that this is indeed the case. To see this, let us consider that the strength of the quantum fluctuations  of the field is not homogeneous in the cavity. If the walls impose Dirichlet boundary conditions, the field modes are sine functions that vanish at the walls, implying that an atom experiences the field's vacuum fluctuations as the weaker the closer the atom is to a wall. Its diamagnetic interaction term should therefore drop in energy as the wall is approached, thus contributing an attractive force. Its paramagnetic term should instead drop in the direction of greater field strength fluctuations, i.e, away from the wall, thus leading to a repulsive force. This is indeed what we found. Further, in a cavity with Neumann boundary conditions, the field's fluctuations are strongest at the walls, implying that then the paramagnetic term should lead to an attractive force and the diamagnetic term to a repulsive force, which is again what we found. 

We also found that depending on whether we include the diamagnetic term in our model or not, the effect of the atom on the cavity walls changes in a manner that is consistent with the interpretation above. Indeed, when the $\phi^2$ term  is not present, the effect is to create a repulsive force between the plates, thus opposing the usual Casimir forces between conducting plates. If we model a dielectric medium by adding more atoms to the cavity then this force can eventually build up to the point of overcoming the regular Casimir force between the conducting plates of an empty cavity. We considered these forces for two different dynamical constraints on the position of the atoms: either as they move along with the plates, or fixed with respect to one of them. With these results, we then built a simple model of a medium between conducting walls that creates a repulsive force between them.
Essentially, this is a simple microscopic account of a setup based on a cavity filled with a medium, for which repulsive forces appear. We then found that when including the diamagnetic term, for which it is necessary to consider a specific spatial shape of the atoms, the forces induced in the plates by the presence of the atoms can become of an attractive kind, which is consistent with the macroscopic approach of Lifshitz. 

In conclusion, we found that the paramagnetic and diamagnetic terms in small system's interaction Hamiltonians tend to contribute opposing Casimir-type forces. The direction of these Casimir forces due to quantum field fluctuations can be predicted in analogy to the direction of the paramagnetic and diamagnetic forces in finite classical background fields. It will be very interesting to extend our analysis to atoms minimally coupled to the electromagnetic field in 3+1 dimensions. 
In this case, the relative strength of the paramagnetic and diamagnetic terms should depend on both the size of the smeared atom and on the details of the shape of the smearing function, beyond the dipole approximation. Namely, the smaller the atom, the larger are the quantum field fluctuations that it couples to and therefore the larger is the diamagnetic term. The detailed shape of the atom's smearing function impacts the strength of the  paramagnetic term through its spatial derivatives. 
Here in the UdW model,  we have shown that the strength of the diamagnetic term, controlled by the parameter $\alpha$  in (\ref{eq:Hi2}) and dependent on the geometry of the problem, is key in order to see the repulsive or attractive nature of the Casimir-Polder forces. In particular,  the choice $\alpha =1$ is  analogous to the 3D model studied in the original Casimir-Polder paper  \cite{CasimirPolder}. We showed that the diamagnetic term can not only contribute significantly but that it can also dominate over the paramagnetic term, thereby determining the direction of the Casimir force. 

It should be very interesting to extend our study to  engineered systems such as metamaterials and analog models for paramagnetic and diamagnetic Hamiltonians such as in superconducting circuits, where various kinds of boundary conditions and coupling strengths can be engineered  \cite{Wallraff,Pozar}. Also, our results suggest that it should be interesting to study the consequences of taking into account the diamagnetic term in those scenarios where the UdW model is traditionally applied without this term, namely, in quantum field theory in curved spacetime (see, e.g., \cite{BandD}). 
$$$$
\section{Acknowledgements} The authors would like to thank Pol Forn-Diaz and Pablo Rodriguez-Lopez for helpful discussions. Achim Kempf acknowledges support from the Canada Research Chairs and Discovery programs of the Canadian Natural Sciences and Engineering Research Council (NSERC). E. Mart\'in-Mart\'inez acknowledges support from the Banting Postdoctoral Fellowship program of NSERC. 
$$$$
\appendix

\begin{widetext}
\section{The dependence of the Casimir-Polder force with the parameter $\alpha$}\label{appfig}

In the main text of the paper, we only consider the two cases when $\alpha=0$ and $\alpha=1$. However, it is interesting to consider how the behaviour of the force changes with the value of $\alpha$. As discussed in section \ref{sec:phisqrd}, $\alpha$ is influenced by the geometry of the atomic systems considered. In this appendix we determine within the UdW model the Casimir-Polder force experienced by an atom in a fixed position of a Dirichlet cavity in the ground state (see Fig. \ref{fig:Alpha}). We find that the force changes sign quickly from repulsive to attractive as $\alpha$ increases and the diamagnetic term starts to dominate over the paramagnetic one, illustrating the effect that the introduction of a paramagnetic term in the interaction term has over the force.

\begin{centering}
\begin{figure}[h]
\includegraphics[width=0.55\textwidth]{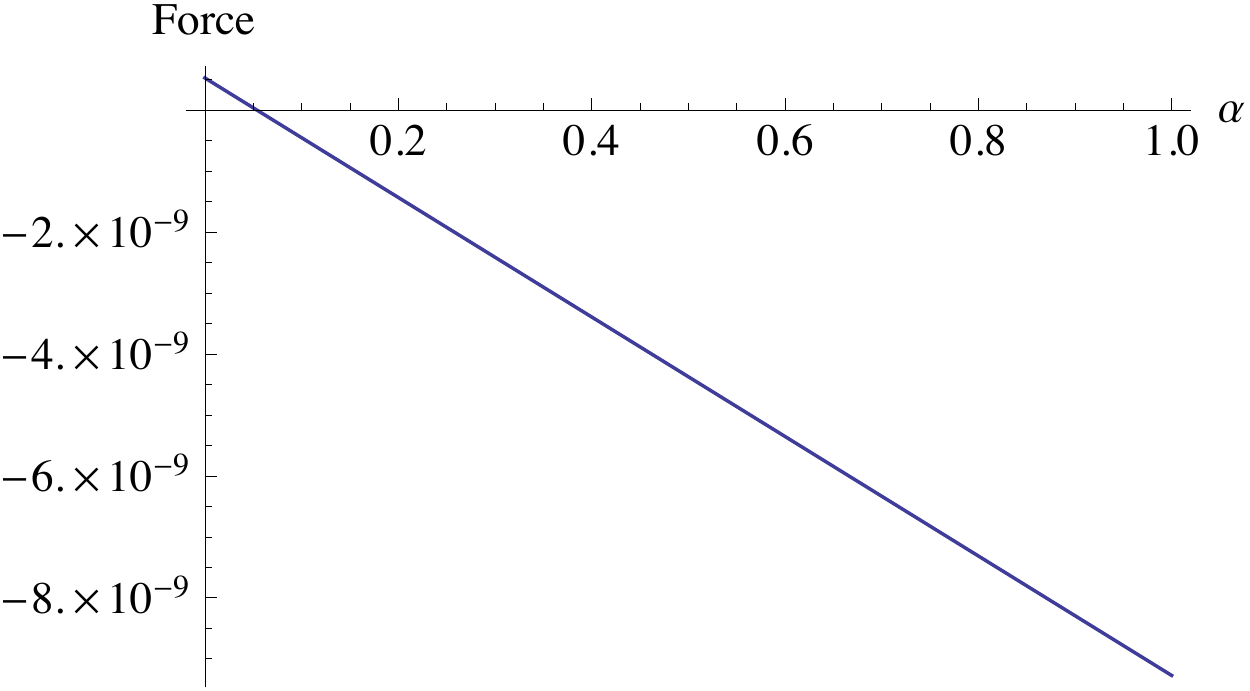}
\caption {Force in the UdW model (in units of $1/L^2$) on an atom at position $x/L=0.1$ as a function of $\alpha$ as defined in\eqref{eq:Hi2}. We chose the parameters $L=1$ and $\lambda=10^{-4}$ (all energies in units of $1/L$). The case $\alpha=0$ (no diamagnetic term) corresponds to Eq. \eqref{preH}. In this case, we see that force repulses the atom from the nearest plate. This behaviour changes quickly to attraction for increasing $\alpha$, and in particular in the case $\alpha=1$, the case which most resembles the full electromagnetic Casimir-Polder case.} \label{fig:Alpha}
\end{figure}
\end{centering}

\section{Analytical expressions for the energy} \label{EnergySI}
In this appendix we show the exact analytical expressions of the energy  of a single-atom situation for the cases in which we include the self-interaction energy and we hence assume a spatial profile for the atom. These have a significantly more complex expression than in the purely UdW case, which are \eqref{eq:secen} for the reflective (Dirichlet) cavity and \eqref{eq:secenVN} for the Neumann-type cavity. 
For the Dirichlet case that energy, which comes from the expression in \eqref{eq:secen2}, is

\begin{align}
E_\text{D}^{(2)}=&\frac{\lambda ^2}{4 \pi ^2 \Omega } \left(\frac{\pi ^2}{\left(a_0^2 \Omega ^2+1\right)^2 e^{\frac{(2 i) \pi  x}{L}}} \left(\frac{a_0 (a_0 \Omega +i)^2 (a_0 \Omega -2 i)}{-L+i \pi  a_0} \left(\, _2F_1\left(1,\frac{i L}{a_0 \pi }+1;\frac{i L}{a_0 \pi }+2;e^{-\frac{(2 i) \pi  x}{L}}\right) 
\right. \right. \right. \\ \nonumber & \left. \left. \left.
+e^{\frac{(4 i) \pi  x}{L}} \, _2F_1\left(1,\frac{i L}{a_0 \pi }+1;\frac{i L}{a_0 \pi }+2;e^{\frac{(2 i) \pi  x}{L}}\right)\right)-\frac{a_0 (a_0 \Omega -i)^2 (a_0 \Omega +2 i)}{L+i \pi  a_0} \left(\, _2F_1\left(1,1-\frac{i L}{a_0 \pi };2-\frac{i L}{a_0 \pi };e^{-\frac{(2 i) \pi  x}{L}}\right)
\right. \right. \right. \\ \nonumber & \left. \left. \left.
+e^{\frac{(4 i) \pi  x}{L}} \, _2F_1\left(1,1-\frac{i L}{a_0 \pi };2-\frac{i L}{a_0 \pi };e^{\frac{(2 i) \pi  x}{L}}\right)\right)-\frac{4 \, _2F_1\left(1,\frac{L \Omega }{\pi }+1;\frac{L \Omega }{\pi }+2;e^{-\frac{(2 i) \pi  x}{L}}\right)}{L \Omega +\pi }
\right. \right.  \\ \nonumber & \left. \left.
-\frac{4 \left(e^{\frac{(2 i) \pi  x}{L}}\right)^{1-\frac{L \Omega }{\pi }} B\left(e^{\frac{(2 i) \pi  x}{L}};\frac{L \Omega }{\pi }+1,0\right)}{\pi }\right)-\frac{8 \pi  \psi ^{(0)}\left(\frac{L \Omega }{\pi }+1\right)}{\left(a_0^2 \Omega ^2+1\right)^2}+\frac{L \Phi \left[e^{-\frac{(2 i) \pi  x}{L}},2,\frac{i L}{a_0 \pi }+1\right]}{a_0 (a_0 \Omega -i) e^{\frac{(2 i) \pi  x}{L}}}
\right.  \\ \nonumber & \left.
+\frac{L \Phi \left[e^{-\frac{(2 i) \pi  x}{L}},2,1-\frac{i L}{a_0 \pi }\right]}{a_0 (a_0 \Omega +i) e^{\frac{(2 i) \pi  x}{L}}}+\frac{L e^{\frac{(2 i) \pi  x}{L}} \Phi \left[e^{\frac{(2 i) \pi  x}{L}},2,\frac{i L}{a_0 \pi }+1\right]}{a_0 (a_0 \Omega -i)}+\frac{L e^{\frac{(2 i) \pi  x}{L}} \Phi \left[e^{\frac{(2 i) \pi  x}{L}},2,1-\frac{i L}{a_0 \pi }\right]}{a_0 (a_0 \Omega +i)}
\right.  \\ \nonumber & \left.
+\frac{\pi  (-4-(2 i) a_0 \Omega ) \psi ^{(0)}\left(\frac{i L}{a_0 \pi }+1\right)}{(a_0 \Omega -i)^2}+\frac{(2 i) \pi  (a_0 \Omega +2 i) \psi ^{(0)}\left(1-\frac{i L}{a_0 \pi }\right)}{(a_0 \Omega +i)^2}-\frac{2 L \psi ^{(1)}\left(\frac{i L}{a_0 \pi }+1\right)}{a_0 (a_0 \Omega -i)}-\frac{2 L \psi ^{(1)}\left(1-\frac{i L}{a_0 \pi }\right)}{a_0 (a_0 \Omega +i)}\right)\end{align}

Similarly, for the energy of a single atom in a Neumann cavity we have the exact anaytical expression, again considering the self-interaction energy and the spatial profile for the atom, as given by \eqref{eq:RegSelfEVN}. That is

\begin{align}
&E_\text{VN}^{(2)}=\frac{\lambda ^2}{4 a_0 \Omega  \left(\pi  a_0^2 \Omega ^2+\pi \right)^2 e^{\frac{(2 i) \pi  x}{L}}} \left(e^{\frac{(2 i) \pi  x}{L}} \left(-(2 i) \pi  a_0 \left(a_0^3 \Omega ^3+3 a_0 \Omega +2 i\right) \psi ^{(0)}\left(\frac{i L}{a_0 \pi }+1\right)
\right. \right.  \\ \nonumber & \left. \left.
+2 \pi  a_0 \left(2+i a_0 \Omega  \left(a_0^2 \Omega ^2+3\right)\right) \psi ^{(0)}\left(1-\frac{i L}{a_0 \pi }\right)+L (a_0 \Omega +i) (a_0 \Omega -i)^2 \left(-e^{\frac{(2 i) \pi  x}{L}}\right) \Phi \left[e^{\frac{(2 i) \pi  x}{L}},2,1-\frac{i L}{a_0 \pi }\right]
\right. \right.  \\ \nonumber & \left. \left.
-L (a_0 \Omega +i)^2 (a_0 \Omega -i) e^{\frac{(2 i) \pi  x}{L}} \Phi \left[e^{\frac{(2 i) \pi  x}{L}},2,\frac{i L}{a_0 \pi }+1\right] 
\right. \right. \\ & \nonumber \left. \left.
+\frac{\pi ^2 a_0}{e^{\frac{(2 i) \pi  x}{L}}} \left(\frac{a_0 (2+i a_0 \Omega ) (a_0 \Omega +i)^2}{\pi  a_0+i L} \left(\, _2F_1\left(1,\frac{i L}{a_0 \pi }+1;\frac{i L}{a_0 \pi }+2;e^{-\frac{(2 i) \pi  x}{L}}\right)
\right. \right. \right. \right. \\ \nonumber & \left. \left. \left. \left.
+e^{\frac{(4 i) \pi  x}{L}} \, _2F_1\left(1,\frac{i L}{a_0 \pi }+1;\frac{i L}{a_0 \pi }+2;e^{\frac{(2 i) \pi  x}{L}}\right)\right)+\frac{a_0 (a_0 \Omega -i)^2 (a_0 \Omega +2 i)}{L+i \pi  a_0} \left(\, _2F_1\left(1,1-\frac{i L}{a_0 \pi };2-\frac{i L}{a_0 \pi };e^{-\frac{(2 i) \pi  x}{L}}\right)
\right. \right. \right. \right. \\ \nonumber & \left. \left. \left. \left.
+e^{\frac{(4 i) \pi  x}{L}} \, _2F_1\left(1,1-\frac{i L}{a_0 \pi };2-\frac{i L}{a_0 \pi };e^{\frac{(2 i) \pi  x}{L}}\right)\right)+\frac{4 \, _2F_1\left(1,\frac{L \Omega }{\pi }+1;\frac{L \Omega }{\pi }+2;e^{-\frac{(2 i) \pi  x}{L}}\right)}{L \Omega +\pi }\right)+\frac{4 \pi  a_0 B\left(e^{\frac{(2 i) \pi  x}{L}};\frac{L \Omega }{\pi }+1,0\right)}{\left(e^{\frac{(2 i) \pi  x}{L}}\right)^{\frac{L \Omega }{\pi }}}
\right. \right. \\ \nonumber & \left. \left.
-2 L (a_0 \Omega +i) (a_0 \Omega -i)^2 \psi ^{(1)}\left(1-\frac{i L}{a_0 \pi }\right)-2 L (a_0 \Omega +i)^2 (a_0 \Omega -i) \psi ^{(1)}\left(\frac{i L}{a_0 \pi }+1\right)-8 \pi  a_0 \psi ^{(0)}\left(\frac{L \Omega }{\pi }+1\right)\right)
\right. \\ \nonumber & \left. 
-L (a_0 \Omega +i) (a_0 \Omega -i)^2 \Phi \left[e^{-\frac{(2 i) \pi  x}{L}},2,1-\frac{i L}{a_0 \pi }\right]-L (a_0 \Omega +i)^2 (a_0 \Omega -i) \Phi \left[e^{-\frac{(2 i) \pi  x}{L}},2,\frac{i L}{a_0 \pi }+1\right]\right)
\end{align}

Where $\Phi\left[z,s,\alpha \right]$ is as specified above in \eqref{eq:definition}, and we further define Gauss' hypergeometric function, 

\begin{align}
_2F_1\left(a,b;c;z\right)=\sum_{n=0}^\infty \frac{(a)_n (b)_n}{(c)_n} \frac{z^n}{n!},
\end{align}
the $n$-th derivative of the digamma function,
\begin{align}
\psi ^{(n)}\left(x\right)=\frac{d^n}{dx^n}\frac{\Gamma'(x)}{\Gamma(x)},
\end{align}
and the incomplete beta function,
\begin{align}
B\left(x;a,b\right)=\int_0^x t^{a-1} (1-t)^{b-1} dt .
\end{align}

Where $\Gamma(x)$ is the gamma function and the Pochhammer symbol is defined as
\begin{align}
(p)_n=\left\lbrace
  \begin{array}{l}
     1 \text{ if } n = 0 \\
     p(p+1)\cdots(q+n-1) \text{ if } n > 0 \\
  \end{array}
  \right.
\end{align}

\section{4th order interaction energy and force between pairs of atoms} \label{4thE}
The energy of interaction of two atoms with positions $x_a$ and $x_b$, without considering the diamagnetic term, and in a Dirichlet cavity reads: 
\begin{align}
&E_{a,b}^{(4)}=\sum_{j=1}^\infty \sum_{l=1}^\infty -\frac{\lambda ^4 L^2}{\pi ^3 j l \Omega  (j+l) (\pi  j+L \Omega )^2 (\pi  l+L \Omega )} 
 \left(2 \left(2 \pi  L \Omega  (j+2 l)+\pi ^2 j (j+l)+2 L^2 \Omega ^2\right) \sin \left(\frac{\pi  j x_a}{L}\right)  \sin \left(\frac{\pi  j x_b}{L}\right) 
 \right. \\ & \left. \nonumber
 \sin \left(\frac{\pi  l x_a}{L}\right) \sin \left(\frac{\pi  l x_b}{L}\right)
 +L \Omega  \sin ^2\left(\frac{\pi  j x_a}{L}\right) \left((\pi  j+3 \pi  l+2 L \Omega ) \sin ^2\left(\frac{\pi  l x_a}{L}\right)+2 \pi  (j+l) \sin ^2\left(\frac{\pi  l x_b}{L}\right)\right)
 \right. \\ & \left. \nonumber
 +L \Omega  \sin ^2\left(\frac{\pi  j x_b}{L}\right) \left(2 \pi  (j+l) \sin ^2\left(\frac{\pi  l x_a}{L}\right)+(\pi  j+3 \pi  l+2 L \Omega ) \sin ^2\left(\frac{\pi  l x_b}{L}\right)\right)\right)
\end{align}
The corresponding force, when the two atoms are kept at a fixed $x/L$, reads:
\begin{align}
&F_{a,b}^{x/L}=-\frac{\text{d}E_{a,b}^{(4)}}{\text{d}L}=\sum_{j=1}^\infty \sum_{l=1}^\infty
\frac{\lambda ^4 L }{\pi ^3 j l \Omega  (j+l) (\pi  j+L \Omega )^3 (\pi  l+L \Omega )^2}
\left(L \Omega  \sin ^2( \frac{ \pi  j x_a}{L}) \left(\sin ^2(  \frac{\pi l x_a}{L} ) \left(\pi ^2 L \Omega  \left(2 j^2+15 j l+3 l^2\right)
\right. \right. \right. \nonumber \\  & \left.  \left. \left.
+2 \pi  L^2 \Omega ^2 (3 j+2 l)+3 \pi ^3 j l (j+3 l)+2 L^3 \Omega ^3\right)+2 \pi ^2 (j+l) \sin ^2(  \frac{\pi l x_b}{L} ) (L \Omega  (2 j+l)+3 \pi  j l)\right)
\right. \nonumber \\  & \left. 
+L \Omega  \sin ^2( \frac{\pi  j x_b}{L}) \left(\sin ^2( \frac{\pi l x_b}{L}) \left(\pi ^2 L \Omega  \left(2 j^2+15 j l+3 l^2\right)+2 \pi  L^2 \Omega ^2 (3 j+2 l)+3 \pi ^3 j l (j+3 l)+2 L^3 \Omega ^3\right)
\right. \right. \nonumber \\  & \left. \left.
+2 \pi ^2 (j+l) \sin ^2(  \frac{\pi l x_a}{L}) (L \Omega  (2 j+l)+3 \pi  j l)\right)+2 \sin (\frac{\pi  j x_a}{L}) \sin (\frac{\pi  j x_b}{L}) \sin (\frac{\pi l x_a}{L} ) \sin (\frac{\pi l x_b}{L}) \left(\pi ^2 L^2 \Omega ^2 \left(3 j^2+17 j l+4 l^2\right)
\right. \right. \nonumber \\  & \left. \left.
+2 \pi ^4 j^2 l (j+l)+2 \pi  L^3 \Omega ^3 (3 j+2 l)+\pi ^3 j L \Omega  (j+3 l) (j+4 l)+2 L^4 \Omega ^4\right)\right) \label{eq:4thF}
\end{align}

An instance of this force as a function of the position of atoms (placed symmetrically) in a cavity is shown in Fig. (\ref{fig:4thF}). 
\end{widetext}

\bibliography{casimir_bib}

\end{document}